\documentclass[fleqn,usenatbib,useAMS]{mnras}
\usepackage{graphicx}
\usepackage{txfonts}
\usepackage{xcolor}

\usepackage[normalem]{ulem}

\definecolor{green}{RGB}{0,150,0}

\usepackage{url}

\newcommand{\logg}{$\log g$}

\newcommand{\simlt}{\lower.5ex\hbox{$\; \buildrel < \over \sim \;$}}
\newcommand{\simgt}{\lower.5ex\hbox{$\; \buildrel > \over \sim \;$}}

\newcommand{\kms}{\,km\,s$^{-1}$}

\newcommand{\Msol}{${\rm M}_{\odot}$}

\newcommand{\Teff}{${\rm T}_{\mbox{\scriptsize eff}}$}
\newcommand{\Feh}{\mbox{$\mbox{[Fe/H]}$}}
\newcommand{\Mh}{\mbox{$\mbox{[M/H]}$}}

\newcommand{\MgFe}{\mbox{$\mbox{[Mg/Fe]}$}}

\newcommand{\aFe}{\mbox{$\mbox{[$\alpha$/Fe]}$}}

\newcommand{\alfa}{$\alpha$}

\def\lesssim{\mathrel{\hbox{\rlap{\hbox{\lower4pt\hbox{$\sim$}}}\hbox{$<$}}}}
\def\gtrsim{\mathrel{\hbox{\rlap{\hbox{\lower4pt\hbox{$\sim$}}}\hbox{$>$}}}}
\def \aj {AJ}
\def \mnras {MNRAS}
\def \apj {ApJ}
\def \apjs {ApJS}
\def \apjl {ApJL}
\def \aap {A\&A}
\def \aaps {A\&AS}

\def \nata {Nature Astronomy}
\def \araa {ARAA}
\def \pasp {PASP}

\usepackage[T1]{fontenc}
\usepackage{ae,aecompl}

\title[Surface Brightness Fluctuation model spectra]
{Surface Brightness Fluctuation spectra to constrain stellar population properties}

\author[Vazdekis et al.]{A. Vazdekis$^{1,2}$\thanks{E-mail:vazdekis@iac.es}
M. Cervi\~no$^{3}$,
M. Montes$^{4}$,
I. Mart\'{\i}n-Navarro$^{1,2,5,6}$,
\newauthor
M. A. Beasley$^{1,2}$
\\
$^{1}$Instituto de Astrof\'\i sica de Canarias (IAC), E-38200 La Laguna, Tenerife, Spain\\
$^{2}$Departamento de Astrof\'\i sica, Universidad de La Laguna, E-38205, Tenerife, Spain\\
$^{3}$Centro de Astrobiolog{\'{\i}}a (CSIC/INTA), ESAC Campus, Camino Bajo del Castillo s/n, E-28692 Villanueva de la Ca\~nada, Spain\\
$^{4}$School of Physics, University of New South Wales, Sydney, NSW 2052, Australia\\
$^{5}$Max-Planck Institut f\"ur Astronomie, Konigstuhl 17, D-69117 Heidelberg, Germany\\
$^{6}$University of California Santa Cruz, 1156 High Street, Santa Cruz, CA 95064, USA
}
\date{}
\pubyear{}
\begin{document}
\label{firstpage}
\pagerange{\pageref{firstpage}--\pageref{lastpage}}
\maketitle
\begin{abstract}

We present a new set of Surface Brightness Fluctuation spectra computed with the
E-MILES stellar population synthesis models. The model SBF spectra cover the
range $\lambda\lambda$ $1680-50000$\,\AA\ at moderately high resolution, all
based on extensive empirical stellar libraries.  The models span the metallicity
range $-2.3\leq\Mh\leq+0.26$ for a suite of IMF types with varying slopes. 
These predictions can complement and aid fluctuation magnitudes studies,
permitting a first order approximation by applying filter responses to the SBF
spectra to obtain spectroscopic SBF magnitudes. We provide a recipe for obtaining
the latter and discuss their uncertainties and limitations. We compare our
spectroscopic SBF magnitudes to photometric data of a sample of early-type
galaxies. We also show that the SBF spectra can be very useful for constraining
relevant stellar population parameters. We find small ($<5$\%) mass-fractions of
extremely metal-poor components ($\Mh<-1$) on the top of the dominant, old and
metal-rich stellar population. These results put stringent constraints on the
early stages of galaxy formation in massive elliptical galaxies. This is
remarkable given the high degree of degeneracy of the standard
spectral analysis to such metal-poor stellar populations in the visible and in
the near-IR. The new SBF models show great potential for exploiting ongoing
surveys, particularly those based on narrow-band filters. 

\end{abstract}



%
\section{Introduction}
\label{sec:intro}

Surface Brightness Fluctuations (SBFs) were first introduced by \cite{TS88} and
\cite{Tonry90} as a way to measure extragalactic distances. These fluctuations
are the result of differences in the luminosity distribution of stars that 
contribute to the flux in each resolution element(s). The SBF is defined as
the variance of these fluctuations, normalized to the mean flux of the galaxy in
each resolution element. The latter is determined locally as the mean of nearby
elements and might require subtracting a smooth galaxy model. Given its
connection with the stellar content, a purely theoretical stellar population SBF
can be defined as the ratio between the second (variance) and first (mean)
moments of the stellar luminosity function that would be obtained for a stellar
population.  The resultant SBF is an intrinsic property of the stellar
population, intimately related to its evolutionary status. Note
that on the top of this signal, there is also a small contribution (less than
$0.1\%$) of Poissonian statistics on the number of stars in different resolution
elements \citep{CLJ08} and, therefore, the SBFs are virtually independent on the
amount of stars in the system.

The observational methodology proposed to derive SBF magnitudes
\citep{TS88} requires high quality photometric data. The Fourier Transform (FT)
analysis is commonly applied in order to disentangle  different signal
contaminations. The FT one to obtain  accurate local means by disentangling
the Point Spread Function (PSF), correlated population variances and the
PSF-uncorrelated noise variance due the observational data acquisition and data
processing. In addition it is also a requirement that each resolution element
contains a large enough number of stars \cite[at least a few tens of
giants][]{TS88} to avoid statistical biases in the determination of the
corresponding moments\footnote{We note that the requirement of having a few
tens of giant stars is equivalent to have the resolution element populated by,
at least, several thousands of stars, leading to gaussian distributions of
integrated luminosities \citep{CLCL06,CLJ08}. Put in other words, in terms of
SBF inferences it implies no correlation between the mean and the variance.}.
The fact that the I band fluctuation magnitude for old and metal rich stellar
populations depends on the galaxy colour with a small scatter has made the SBF
technique a popular secondary distance calibrator (e.g., \citealt{Jacetal92}).
In fact SBFs provide robust empirical distance calibrations, such as those of
\citet{Blakeslee10} and \citet{Cantiello18}.

The comparison of theoretical SBF predictions with integrated fluctuation
magnitudes and colours is potentially able to provide additional constraints on
stellar populations (e.g., \citealt{Worthey94}). It has been shown that
the use of the SBFs is able to break the age-metallicity degeneracy affecting
old stellar populations \citep{Worthey94,Cantiello03}. Their relatively modest
development can be attributed to a great extent to the difficulties in obtaining
SBF magnitudes in more than a single band for the same galaxy. However, there are
examples in the literature that have made use of SBF magnitudes and colours and
integrated colours to study the stellar populations via comparison with model
predictions (e.g.,
\citealt{Blakeslee01,Cantiello07,Cantiello11,Jensen01,Jensen03,Liu00,Liu02}),
including galaxy stellar population gradients (e.g.,
\citealt{Cantiello05,Jensen15}). Such analyses have also been performed in
globular clusters (e.g., \citealt{MFA06}), including the relation between
stellar evolutionary modelling and SBF predictions in the Magellanic Clouds
\citep*{GLB04,Mouhcine05,Raimondo09}.

Near-IR SBFs have also been used to constrain the properties of stellar populations \citep{Liu00,GLB04} and to test different mass-loss rates affecting the evolution of Thermally Pulsing Asymptotic Giant Branch (TP-AGB) stars. In fact, \citet{Raietal05} showed that NIR SBFs can be used to disentangle observable properties of TP-AGBs. This is particularly relevant for galaxies that contain significant fractions of stellar populations of intermediate ages, which are heavily contributed by these stars. Given the strong connection between the integrated light and its stars,  SBFs have been used to test the impact of stellar winds in the evolution of AGB and TP-AGB stars taking advantage of the large sensitivity of of the SBFs to these stars, particularly in the IR spectral range \citep{GLL10,GLL18}. Moreover, SBFs have been also used to determine the faint end of the galaxy number counts in the Hubble Deep Field-North, extending it by $2$\,mag beyond the limits of photometric studies \citep{MFA03}. 

Despite the fact that most SBF studies have been performed at the photometric level, the method can be also used in spectroscopy. \cite{Buzz93} modeled the first low resolution SBF spectrum (see their Figure 1). These models, as well as their predictions shown in \cite{Buzz89}, are tabulated in terms of their effective number of stars, $\cal N$\footnote{$\cal N$ is defined as the ratio between the square of the mean over the variance of the stellar luminosity distribution. Therefore the product of $\cal N$ and SBF gives the total flux (see \citealt{Buzz89} and \citealt{Buzz93} Sect.~\ref{sec:others})}. The SBF spectrum has shown its ability to break stellar population degeneracies by employing particular spectral features in high resolution spectra \cite[see, e.g.,][their Fig.~9 and related discussion]{Buzz05}.
Additional SBF spectra, again tabulated as $\cal N$, were provided by \cite{GDetal05} and presented in \cite{CLVallarta09} and \cite{Cer13} within the context of metric fitting for stellar population studies.

The first observational SBF spectrum of a galaxy was recently presented in \citet{Mitzkus18}. These authors employed the FT analysis as proposed by \cite{TS88} to obtain the SBF spectrum of a nearby S0 galaxy, NGC~5102, using data from MUSE Integral Field Spectroscopic (IFS) instrument. These authors also computed model SBF spectra to compare their derived SBF spectrum. They found that including the SBF spectra in the analysis lead to additional constraints in relevant stellar population parameters. Although the pioneering work of these authors has opened a possible way to use IFS data to obtain SBF galaxy spectra, the methodology is currently in its infancy and new investigations are badly needed. 

SBF spectral models provide the possibility to derive robust distance estimates, measure SBF magnitudes and constrain stellar populations in galaxies. Moreover, formally, the SBF spectrum should not be restricted necessarily to a particular object, but the method could be extended to any galaxy spectral sample where a common mean and a variance can be estimated.

Here we present new model SBF spectra at moderately high resolution covering a wide range of ages, metallicities and varying IMFs. The main model ingredients are described in Section~\ref{sec:ingredients} while the computational details are provided in Section~\ref{sec:SBF_spectra}. The latter also includes a description of the coverage and behaviour of these models as a function of relevant stellar population parameters. Section~\ref{sec:use} provides useful details on how the newly computed SBF spectra can be used for several applications. We compare these new SBF spectra with those predicted by other authors in the literature in Section~\ref{sec:others}. Finally, we show potential useful applications of these models in Section~\ref{sec:applications} and summarize our results in Section~\ref{sec:summary}.

\section{Model ingredients}
\label{sec:ingredients}

\subsection{Isochrones}
\label{sec:isochrones}

We employ  two sets of solar-scaled theoretical isochrones of
\citet{Padova00} (hereafter Padova00) and \citet{Pietrinferni04} (hereafter
BaSTI). The Padova00 isochrones cover a wide range of ages, from $0.063$ to
$17.8$\,Gyr, and six metallicity bins, where $Z=0.019$ represents the solar value. The range of
initial stellar masses extends from $0.15$ to $7$\,\Msol. A helium fraction was
adopted according to the relation: $Y\approx0.23+2.25Z$. The empirical relation by \citet{Reimers} is adopted for the mass-loss rates, multiplied by a parameter $\eta$ that is set to $0.4$ \citep{RenziniFusiPessi88}.

The BaSTI theoretical isochrones of \citet{Pietrinferni04} were extended as described in \citet{Vazdekis15}, including an
extra (supersolar) metallicity bin, and extending the isochrones to the very
low-mass (VLM) regime down to $0.1$\,\Msol, based on the models of
\citet{Cassisi00}. We note that the temperatures for these stars are cooler than
those in Padova00 \citep{MIUSCATI}. A complete description of the BaSTI database
can be found in 
\citet{Pietrinferni04,Pietrinferni06,Pietrinferni09,Pietrinferni13} and
\citet{Cordier07}. We adopted the non-canonical BaSTI models with the mass loss
efficiency of the Reimers law \citep{Reimers} set to $\eta=0.4$. The initial He
mass fraction ranges from $0.245$ to $0.303$, for the more metal-poor to the
more metal-rich composition, respectively, with $\Delta Y / \Delta Z \approx
1.4$. 

The main sequence (MS) loci are in good agreement between these two set of models. The turn-off (TO) stars
are also in good agreement for old stellar populations, but their luminosities
differ for young and intermediate age regimes. The BaSTI isochrones show
systematically cooler red giant branch (RGB) stars at low metallicities,  but a hotter RGB in the high metallicity regime. Due to differences in the adopted mass-loss efficiency
along the RGB, the core He-burning stage is hotter in the BaSTI models. Conversely, this situation reverses for young ages with the Padova00 models showing more extended blue loops, due to the treatment of convection of the He-burning core in intermediate-mass stars. Both sets of models include the TP-AGB regime using simple synthetic prescriptions. We note that the synthetic-AGB treatment in the BaSTI isochrones account for AGB nucleosynthesis, i.e. the effects of the third
dredge-up and hot-bottom burning, including the related evolutionary effects. In particular the effect of the change in the
envelope C/O ratio induced by the third dredge-up was mimicked
by adopting a radiative opacity at a constant heavy elements distribution, but
allowing the global metallicity to change. It has been shown that the population synthesis models based on the BaSTI AGB-extended models provide integrated colours that match the observations of super star clusters with intermediate ages \citep{Noel13}. We refer the interested reader to \citet{Cassisietal04}, \citet{Pietrinferni04} and \citet{Vazdekis15} for a more detailed comparison of these two sets of isochrones.

The theoretical parameters of these isochrones are transformed to obtain stellar
fluxes using empirical relations between colours and stellar parameters (\Teff,
$\log g$, \Feh). We use the metallicity-dependent empirical relations of
\citet*{Alonso96} and \citet*{Alonso99} for dwarfs and giants, respectively to
obtain stellar fluxes. These relations link the colours to the stellar
parameters (\Teff, $\log g$, \Feh) based on two extensive photometric stellar
libraries of dwarfs and giants (around $\sim500$ stars each library). We use the
empirical compilation of \citet*{Lejeune97, Lejeune98} (and references therein)
for the coolest dwarfs (\Teff$\lesssim4000$\,K) and giants
(\Teff$\lesssim3500$\,K) for solar metallicity, and also for stars with
temperatures above $\sim8000$\,K. A semi-empirical approach for the low
temperature stars at other metallicities is used by combining these relations
and the model atmosphere predictions of \citet{Bessell89,Bessell91} and the
library of \citet{Fluks94}. Finally, we also use the metal-dependent bolometric
corrections of \citet*{Alonso95} and \citet{Alonso99} for dwarfs and giants,
respectively. We adopt $BC_{\odot}=-0.12$. Assuming $V_\odot=26.75$
\citep{Hayes85} we obtain for the sun the absolute magnitude ${\rm
M_{V_\odot}=4.82}$ and ${\rm M_{{bol}_{\odot}}}$ is given by ${\rm
M_{V_{\odot}}+BC_{V_{\odot}}}= 4.70$.

\subsection{Stellar spectral libraries}
\label{sec:stellar_libraries}
To compute both the SBF and SSP spectra at moderately high resolution we employ a variety of  extensive empirical stellar spectral libraries depending on the wavelength range. We refer the interested reader to \citet{Vazdekis16} for a complete description of the model construction which we briefly summarize here. We employ the MILES library \citep{MILESI}, with the atmospheric parameters of \citet{MILESII}, to build up our reference models in the optical range \citep{MILESIII}. The NGSL library \citep{NGSL}, as fully characterized in \citet{NGSLI}, feed our models in the UV range. To extend our model predictions out to 5\,$\mu$m we employ the Indo-US \citep{Valdes04}, CaT \citep{CATI,CATII} and IRTF \citep{IRTFI,IRTFII} stellar libraries, as described in \citep{MIUSCATI}, \citet{Roeck15} and \citet{Roeck16}. 
In the context of the SBFs it is particularly important to emphasize here that
the IRTF library that we use for the predictions in the near-IR covers all the relevant phases of the more evolved stars such as the AGBs, including five Carbon Stars. However this library has a rather limited metallicity coverage spanning around solar metallicity. These two aspects, together with the reduced number of stars ($\sim180$) that compose the IRTF library, limit the quality of our model predictions in the near-IR spectral range, particularly for intermediate-age regimes and for metallicities outside the range $(-0.4,+0.2)$ (see for details \citealt{Roeck15,Vazdekis15}).

All the atmospheric parameters of these libraries have been homogenized, and placed on the system of \citet{MILESII}. However not
all the stars were used for computing the models as we
checked every single star spectrum and discarded those with
peculiarities, e.g., signs of spectroscopic binaries. In the less severe cases we just decreased their contributing weight within the interpolating algorithm. We refer the reader to the above papers for the full details.

\subsection{IMF shapes}
\label{sec:IMF}

For the IMF, $\Phi(m) = \mathrm{d}N/\mathrm{d}m$, we adopt the multi-part power-law IMFs of \citet{Kroupa01}, i.e.
universal and revised, the two power-law IMFs described in \citet{Vazdekis96},
often regarded as ``unimodal" and ``bimodal", both characterised by the
logarithmic slope ($\mathrm{d}N/\mathrm{d}(\log m)$), $\Gamma$ and $\Gamma_b$, respectively, and the \citet{Chabrier} single-stars IMF. Note that the \citet{Salpeter55} IMF is obtained by adopting the unimodal IMF with $\Gamma=1.35$, and the  Kroupa Universal IMF is very similar  (although not identical) to a bimodal IMF with slope $\Gamma_b=1.3$.  
\section{Model SBF spectra computation}
\label{sec:SBF_spectra}

A single-age, single-metallicity stellar population, SSP, can be understood as a probability distribution that is mainly characterized by a mean $L_{\lambda}^\mathrm{SSPmean}$ and a variance $L_{\lambda}^\mathrm{SSPvar}$ \cite[see][ for details]{CLCL06}. However it is worth noting that when computing SSPs it is a common practice to provide only the mean spectrum of the population. Therefore in its traditional use the SSP is regarded as the mean value of the distribution and it can be identified as    

\begin{equation}
L_{\lambda}^\mathrm{SSP} \equiv L_{\lambda}^\mathrm{SSPmean}
\label{eq:SSPmean}
\end{equation}

Here we follow the "base" modelling approach described in \citet{Vazdekis15} to compute the SBF spectrum corresponding to the SSP. To obtain the mean SSP spectrum, $L_{\lambda}^\mathrm{SSP_{mean}}$, we basically integrate the stellar spectra along the isochrone of the various empirical libraries described in Section\,\ref{sec:stellar_libraries}. The resulting E-MILES $L_{\lambda}^\mathrm{SSP_{mean}}$ model is then used as the reference spectrum over which we calculate the corresponding variance spectrum. We compute $L_{\lambda}^\mathrm{SSP_{mean}}$ as follows

\begin{eqnarray}
L_{\lambda}^\mathrm{SSP_{mean}} &=& \int_{m_{\rm l}}^{m_{\rm t}}
S_{\lambda_{\lambda_\mathrm{ref}}}(m,t,\Feh){F_{\lambda_\mathrm{ref}}}(m,t,\Feh)N_{\Phi}(m,t)\,  \mathrm{d}m \nonumber\\
&=& \int_{m_{\rm l}}^{m_{\rm t}}
L_\lambda(m,t,\Feh) \,N_{\Phi}(m,t)\,\, \mathrm{d}m
\label{eq:SSP}
\end{eqnarray}

\noindent where $S_{\lambda_{\lambda_\mathrm{ref}}}(m,t,\Feh)$ is the empirical
stellar spectrum, normalised in a reference wavelength interval
$\lambda_\mathrm{ref}$, corresponding to a star of mass $m$ and measured
metallicity \Feh, which is alive at the age $t$ of the SSP.
$F_{\lambda_\mathrm{ref}}(m,t,\Feh)$ is the flux of the star in the reference
wavelength interval. Hence $L_\lambda(m,t,\Feh) =
S_{\lambda_{\lambda_\mathrm{ref}}}(m,t,\Feh){F_{\lambda_\mathrm{ref}}}(m,t,\Feh)$
is the luminosity assigned to each star for their given evolutionary parameters
$m$, $t$ and $\Feh$. Finally, $N_{\Phi}(m,t)$ is the probability that the system
contains a star with a given mass.  Note that this value is given by the IMF for
the SSP with a given age, but it could also refer to the case of complex Star
Formation Histories (SFHs).  
The units of the mean SSP spectra,
$L_{\lambda}^\mathrm{SSP_{mean}}$, are $\mathrm{erg}\, \mathrm{s}^{-1}\,
\mathrm{\AA}^{-1}$. We apply a mass normalization by the mean mass
of a star as given by the IMF, $\left< m \right>$, to provide units of
$\mathrm{erg}\, \mathrm{s}^{-1}\, \mathrm{\AA}^{-1}\,\mathrm{M}_\odot^{-1}$. Note that such {\it a posteriori} normalization allows us to scale the obtained luminosity to that of a stellar system with any given mass.

To calculate the SBF spectrum, $L^{\mathrm{SSPsbf}}_{\lambda}(t,\Mh,\Phi)$, of
the SSP with age $t$, total metallicity $\Mh$, and IMF $\Phi$, we need both the
mean $L_{\lambda}^\mathrm{SSPmean}$ and the variance
$L_{\lambda}^\mathrm{SSPvar}$ spectra of the SSP. The variance spectrum, also
regarded as $\sigma^2$, i.e. the square of the standard deviation\footnote{We
recall that in this context, the standard deviation is just a measure of the
variance, but it does not assume any hypothesis about the underlying
distribution. As it is, this standard deviation cannot be used to establish
confidence intervals unless the underlying distribution is proven to be
Gaussian.} $\sigma$ around the mean that is taken as reference spectrum

\begin{equation}
L^{\mathrm{SSPvar}}_{\lambda} = 
\int_{m_{\rm l}}^{m_{\rm t}} L^2_\lambda(m,t,\Feh) \,N_{\Phi}(m,t)\,\, \mathrm{d}m  - ( L^\mathrm{SSPmean}_{\lambda})^2
\label{eq:Fvar}
\end{equation}

\noindent The variance units are $\mathrm{erg}^2\, \mathrm{s}^{-2}\,
\mathrm{\AA}^{-2}$, i.e. the same ones as $( L^\mathrm{SSPmean}_{\lambda})^2$
before the mass normalization (otherwise the right hand side terms of
Eq.~\ref{eq:Fvar} would have different units). Once again, if divided by the
mean mass of a star as given by the IMF we obtain $\mathrm{erg}^2\,
\mathrm{s}^{-2}\, \mathrm{\AA}^{-2}\,\mathrm{M}_\odot^{-1}$ \cite[note that the
correct mass normalization of the variance is $\mathrm{M}_\odot^{-1}$ and not
$\mathrm{M}_\odot^{-2}$; we refer to][for an in-depth
discussion]{CLCL06,Cer13}.  The subtracting term $(
L^\mathrm{SSPmean}_{\lambda})^2$ in Eq.~\ref{eq:Fvar} is the variance of the SSP
as shown in \citet{CLCL06} and \citet{CLJ08}. Note, however, that this term is
usually omitted in SBF modeling works under the assumption of Poissonian
statistics in the number of stars along different element resolutions. 
However, this term only accounts for the $\sim$0.1\% of the final fluctuation
\cite[see][for details and further discussion]{CLJ08}.  On the other hand, we
note that this term is included in the observational approach as shown in
Section~\ref{sec:intro} and Appendix~\ref{sec:galaxy_sbf_spectra}. Note
that the subtracting term must be applied before any mass normalization of
$L^\mathrm{SSPmean}_{\lambda}$.

Finally, the SBF spectrum is computed as the ratio between the variance and the
mean SSP spectra

\begin{eqnarray}
L^{\mathrm{SSPsbf}}_{\lambda}(t,\Mh,\Phi) =
\frac{L^{\mathrm{SSPvar}}_{\lambda}(t,\Mh,\Phi)}{L^{\mathrm{SSPmean}}_{\lambda}(t,\Mh,\Phi)}
\label{eq:SBF}
\end{eqnarray}

\noindent Irrespective of whether
$L^{\mathrm{SSPsbf}}_{\lambda}(t,\Mh,\Phi)$ is computed with or without applying
any mass normalization, the fluctuation spectra units are $\mathrm{erg}\,
\mathrm{s}^{-1}\, \mathrm{\AA}^{-1}$. It is advantageous to provide
separately the variance spectrum associated to the mean SSP spectrum, instead of
providing simply the SBF spectrum. Among other applications, it allows one to
calculate the variance of a complex SFH by a direct integration of the variance
associated with each individual SSP in the SFH, in a similar way as is performed
for the mean flux (see Section~\ref{sec:SFHs}). We stress that the variance
properties are not shared by the SBF as we show below.

In the integration of both the mean and variance, each stellar spectrum is also 
characterized by its \Teff\ and \logg\ parameters, and we scale each 
spectrum according to the flux in the broad-band $V$ filter. This flux is derived 
following the same empirical photometric relations applied to the theoretical 
isochrones. Before scaling we first normalise each stellar spectrum by 
convolving it with the filter response of
\citet{BuserKurucz78}. We follow the method described in \citet{FalconBarroso11}
that is based on the calibration of \citet*{Fukugita95} to assign the absolute
flux to the $V$-band. The zero-point is established by the Vega spectrum of
\citet{Hayes85} with a flux of $3.44\times10^{-9}$\,erg
cm$^{-2}$\,s$^{-1}$\,\AA$^{-1}$ at 5556\,\AA, and the $V$ magnitude is set to
$0.03$\,mag, which is consistent with \citet{Alonso95}. Finally, for the spectral
range beyond $\sim9000$\,\AA, we use as a reference the $K$ band as described in
\citet{Roeck15}. 

We apply a local interpolation scheme as described in \citet{CATIV} (see their
Appendix B), and updated in \citet{Vazdekis15}, to calculate a stellar spectrum
for a given set of atmospheric parameters. This algorithm is particularly
suitable to overcome the gaps and asymmetries present in the parametric
distribution of any empirical library. Note that
a requested stellar spectrum is calculated according to
the stellar parameters (\Teff, \logg, \Feh), irrespective of the evolutionary
stage. It has been shown that such prescription might not be sufficient to model the behaviour of the CO feature as a function of temperature in the $K$ band, particularly for cool AGB stars \citep{Marmol-Queralto08}.  This limitation might have a non-negligible effect on the SBF spectra of intermediate-ages, particularly at  red wavelengths. 

We remark here that our models employ both solar-scaled isochrones and empirical
stellar libraries. Therefore these models, for which we assume that $\Mh=\Feh$,
can be considered "base" models following the description provided in
\citet{Vazdekis15}. Because the library spectra follow the Milky-Way abundance
pattern with metallicity, the computed models are nearly consistent and
solar-scaled around solar metallicity. At subsolar metallicities they lack
consistency as these models combine solar-scaled isochrones with \alfa-enhanced
star spectra. We refer the interested reader to  \citet{Vazdekis15} for a
version of the SSP spectra that is self-consistent in the optical range. 
Hereafter we use \Feh\ for the metallicity of the stars of the empirical
libraries feeding the models as these are measured iron abundance values.
However as the models are computed on the basis of the total metallcity we use
\Mh\ for the resulting predictions, which are identical only in the case
of the solar-scaled assumption implicit in the base modelling approach.

The model spectra in the various spectral ranges, which are computed in a fully
consistent manner with the same population synthesis code, are joined as
described in detail in \citet{Vazdekis16}. To summarize, we identify spectral
regions where no major features are found for the range of ages and
metallicities covered by our models. The overlapping windows have been chosen to
be sufficiently wide to reach enough statistics for the continuum counts and, at
the same time, to avoid the presence of strong spectral features. Finally, we
re-scale, using the selected windows, the spectra blueward and redward of the
MILES range to match the continuum of the different models based on the MILES
library, which is taken as reference. Note that this approach is possible due to
the good flux calibration quality of the stellar spectra of the various
libraries. We apply this joining  strategy to obtain both the E-MILES SSP
spectra and the  corresponding variance spectra. Finally, we divide the two as
previously mentioned in Eq.~\ref{eq:SBF} to obtain the E-MILES SBF spectra.

  \begin{figure}
   \includegraphics[width=0.49\textwidth]{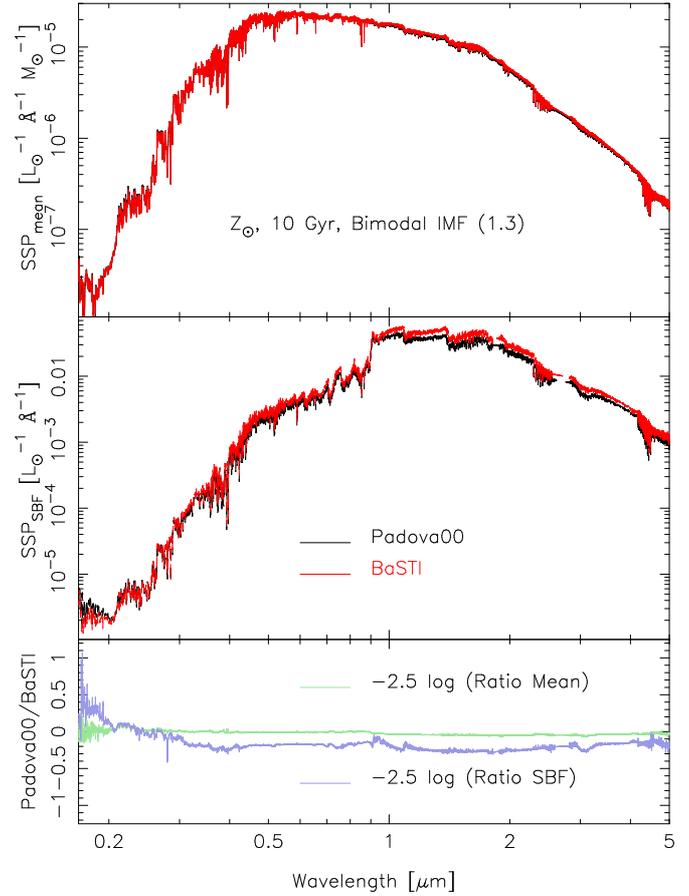}
    \caption{E-MILES SSP (top panel) and SBF (middle panel) spectra of a stellar population of $10$\,Gyr, solar metallicity and standard IMF (bimodal shape with logarithmic slope 1.3). We show the spectra computed with BaSTI (black) and Padova00 (red) isochrones. In the bottom panel we show the ratio of the models computed with different isochrones for the two, the mean SSP spectra (green) and the SBF spectra (blue).}
    \label{fig:sbf_ssp_spec}%
   \end{figure}

Figure~\ref{fig:sbf_ssp_spec} (middle panel) shows the SBF spectrum of a stellar population with solar metallicity and $10$\,Gyr computed with the BaSTI and Padova00 evolutionary models (see Sect.~\ref{sec:isochrones}). For comparison, the figure shows in the upper panel the corresponding SSP spectrum. Note that, given the construction of the SBF spectrum, some features of the SSP get amplified. For example, we see that the drop in flux toward the UV spectral range is far more severe in the SBF spectrum than in the SSP spectrum. However we also recall that the employed isochrones do not include the white dwarfs sequence and, therefore, the UV contribution is dominated by hotter stars in the MS and also in the Horizontal Branch (HB). The latter is only relevant for metal-poor stellar populations, while for the plotted SBF spectrum (old metal-rich) the main contribution in the UV comes from the upper-MS. However these evolutionary phases are statistically far more populated, covering a relatively small range of luminosities. Another important difference is that whereas the SSP spectrum peaks around the $V$ band, the SBF spectrum peaks around the $J$ or $H$ bands. This arises  because in the SBF spectra the contributions from red luminous evolved stars are emphasized with respect to other phases.

It is precisely for this reason that the SBF spectrum also shows significant molecular bands in comparison to the SSP spectrum. This is clearly illustrated in the spectral region covered by the prominent Ca${\rm\sc{II}}$ triplet feature in the SSP spectrum at around $\sim8500$\,\AA, which is almost completely buried by the surrounding molecular bands in the SBF spectrum.
Finally, as expected, we see that in this age and metallicity regime the choice of evolutionary models, Padova00 and BaSTI, has a significantly larger impact on the synthesized SBF spectra in comparison to the SSP spectra (see also Sect.~\ref{subsubsec:evolmod_comp}).

\subsection{Optical SBF model spectra with varying \MgFe-abundance}
\label{sec:alpha}

We also compute a set of models with varying
$\MgFe$\ abundance ratio for the MILES spectral range to assess the effect of
this parameter on the resulting optical SBF spectra. For this purpose we use the
\MgFe-enhanced and solar-scaled version of our models that are described in
\citet{Vazdekis15}. Briefly, the models employ the theoretical stellar
spectral library of \citet{Coelho05}, with its extension to cool stars of
\citet{Coelho07}, to obtain SSP-based differential spectral corrections. For
this purpose the models employ the BaSTI solar-scaled and \alfa-enhanced
isochrones of \citet{Pietrinferni04,Pietrinferni06}, which are calculated with
similar \aFe\ abundance ratios to the stellar atmospheres. The isochrones are
converted to the observational plane using the photometric stellar libraries
described in Section~\ref{sec:isochrones}. We take into account the \MgFe\
determinations for the MILES stars of \citet{Milone11} to compute reference
models based exclusively on the MILES database. Note that we use these \MgFe\
values as a proxy for the \aFe\ values of the MILES stars. For this reason we
refer to the resulting models as \MgFe-enhanced instead of \aFe-enhanced.
Finally we correct these reference SSP models with the corresponding SSP-based
theoretical spectral responses to obtain the enhanced SSP models.

It is worth recalling here that these models are self-consistent in the sense
that both ingredients, the theoretical stellar spectra and the isochrones are
computed with the same overall \alfa-enhancement. These models are computed for
two $\MgFe$-enhancement values: 0.0 and 0.4. The $\MgFe$-enhanced models have an
iron content that differs by $\Delta\Feh=-0.3$ with respect to the total
metallicity \Mh\ (see \citealt{Vazdekis15}). For example, the $\MgFe$-enhanced
SBF model spectra with $\Mh=+0.06$ have $\Feh=-0.24$, whereas the solar-scaled
spectra verify  
$\Mh\equiv\Feh$.

\subsection{Models coverage}
\label{sec:coverage}

The newly computed model spectra cover the wavelength range $\lambda\lambda$ 1680.2\,\AA--49999.4\,\AA\ at moderately high spectral resolution. As the models employ varying stellar spectral libraries, we have decided to
keep their nominal resolutions in most of the spectral ranges. This includes a change from a constant FWHM resolution to a constant $\sigma$ resolution at $8950$\,\AA. Note for example in Fig.~\ref{fig:sbf_ssp_spec} the change in resolution from $5$ to
$2.51$\,\AA\ (FWHM) blueward $3541.4$\,\AA, or that from FWHM $=2.5$\,\AA\ to $\sigma=60$\kms\ redward $8950$\,\AA\ (FWHM $=4.2$\AA\ at that wavelength). The spectra have been rebinned to a linear dispersion of $0.9$\,\AA/pix along the whole spectral range. 

The E-MILES SBF spectra span a rather wide age interval, including ages below
$1$\,Gyr. However the SBF predictions for ages below $3$\,Gyr are
far more sensitive to the modelling details as in, e.g., the mass-loss
along the AGB \citep{Raietal05}; see also \cite{GLL10,GLL18}. Regarding the metallicity coverage the
SBF spectra computed with the Padova00 models are provided for the following
metallicity bins $Z=0.0004$, $0.001$, $0.004$, $0.008$, $0.019$ and $0.03$, or,
alternatively, $\Mh=-2.32, -1.71, -1.31, -0.71, -0.40, 0.0$ and $0.22$,
respectively. On the other hand the models based on BaSTI are provided for the
following metallicity bins: $Z=0.0001$, $0.0003$, $0.0006$, $0.001$, $0.002$,
$0.004$, $0.008$, $0.0100$, $0.0198$, $0.0240$, $0.0300$ and $0.0400$, or,
alternatively, $\Mh=-2.27, -1.79, -1.49, -1.26, -0.96, -0.66, -0.35, -0.25,
0.06, 0.15$ and $0.26$, respectively. With respect to the IMF shape, apart from
the Chabrier and the two Kroupa IMFs, we vary the slope of the unimodal and
bimodal IMF functional forms. Specifically, we compute SBF spectra for top-heavy IMF slopes ($0.3$) to very bottom-heavy ($3.5$). 

The quality of the computed models depend to a great extent on the input stellar libraries, which vary in their atmospheric coverage and density of stars. We refer the interested reader to \citet{Vazdekis16} for a full characterization of the quality in the various spectral ranges covered by E-MILES models. All these models can be downloaded from the MILES website \url{http://miles.iac.es}.


   \begin{figure}
    \includegraphics[width=0.49\textwidth]{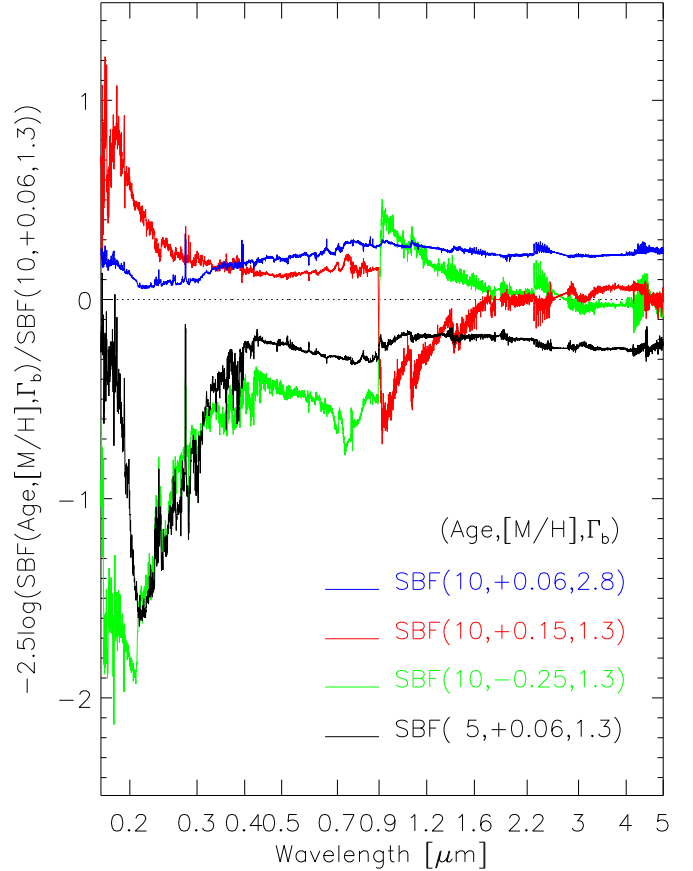}
    \caption{The sensitivity of the SBF spectra to relevant stellar population parameters is illustrated as a difference in magnitude. SBF spectra with varying metallicity (\Mh$=0.15$ and \Mh$=-0.25$ vs. \Mh$=0.06$ in red and green, respectively) for the same age ($10$\,Gyr) and IMF (bimodal with logarithmic slope $\Gamma_b=1.3$). SBF spectra with varying age ($5$ vs. $10$\,Gyr), same metallicity (\Mh$=0.06$) and IMF slope ($\Gamma_{b}=1.3$) is shown in black. Finally SBF spectra with varying IMF, bottom-heavy ($\Gamma_{b}=2.8$), which is characteristic of massive ETGs, and standard ($\Gamma_{b}=1.3$), with the same age ($10$\,Gyr) and metallicity (\Mh$=0.06$) is shown in blue. Note that each of the varying models, whose parameters are indicated within parenthesis (age, metallicity and IMF slope) are compared allways to the same reference SBF spectrum ($10, +0.06, 1.3$). All the models are computed with the BaSTI isochrones.}
    \label{fig:behaviour}%
   \end{figure}

   \begin{figure}
    \includegraphics[width=0.49\textwidth]{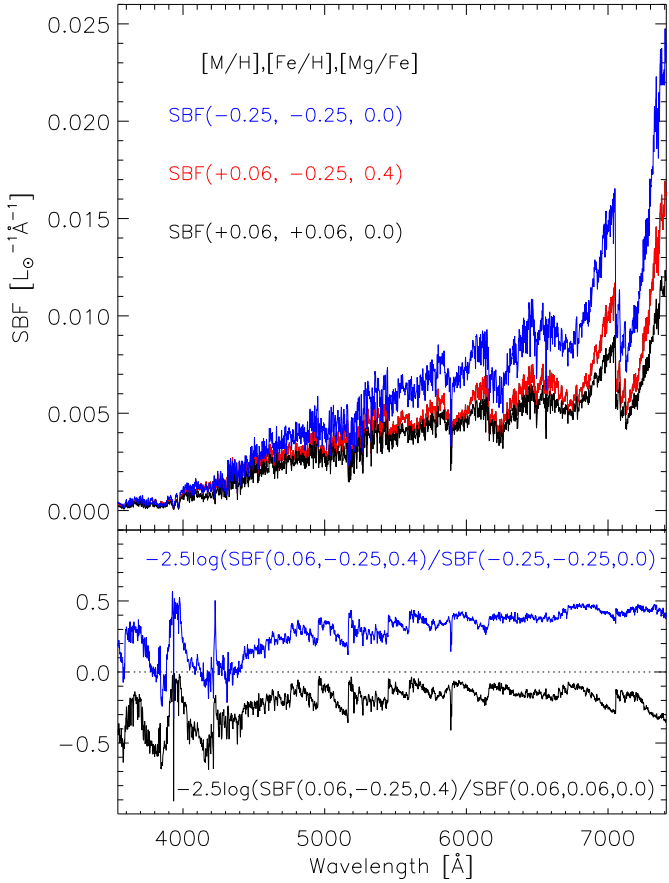}
    \caption{The upper panel shows SBF spectra with constant total metallicity ($\Mh=0.06$) and varying abundance ($\MgFe=0.0$ plotted in black and $\MgFe=0.4$ plotted in red). These two models differ in their \Feh\ content ($\Feh=0.06,-0.25)$. Also plotted (in blue) is a solar-scaled model with similar iron abundance (\Feh$ =-0.25$) but differing total metallicity ($\Mh=-0.25$) with respect to the $\alpha$-enhanced model. All the spectra are computed with the BaSTI models and have similar age ($10$\,Gyr) and IMF (bimodal with logarithmic slope $\Gamma_b=1.3$). The sensitivity of these two solar-scaled SBF spectra with respect to the $\alpha$-enhanced model is shown in the lower panel as a difference in magnitude.}
    \label{fig:alpha}%
   \end{figure}

\subsection{Behaviour of the models}
\label{sec:behaviour}

\subsubsection{Variation with stellar population parameters}

In this section we describe the behaviour of the SBF spectra with relevant
stellar population parameters. Figure~\ref{fig:behaviour} shows the difference in magnitude obtained for SBF spectra with varying metallicity, age and IMF. We see that the SBF spectra are mostly sensitive to the
metallicity and age, particularly in the UV and optical spectral ranges. This variation in the blue spectral range is not surprising since the TO (hot) stars are positioned in the blue part of the HR diagram. The sensitivity to age and metallicity is roughly similar in shape, although some differences can be
seen in the bluest end. On the contrary, the SBF spectral sensitivity obtained when the metallicity is varied shows larger differences in the near-IR, mostly related to the relatively stronger molecular bands and steeper SBF spectrum shape of the more metal-rich model. More importantly, such differences indicate potential capabilities at breaking the age metallicity degeneracy using the SBF spectra \cite[see also][]{Buzz05}. 

Figure~\ref{fig:behaviour} shows that the IMF remains as a second order effect on
the SBF spectra. This is not surprising for such an old age regime as in the computation of an SBF spectrum the relative contribution of the brightest, more evolved, stars is emphasized. As these stars span a rather narrow initial
mass range there is not enough contrast to discern the contribution of lower mass stars, whose contribution has decreased in the computation of the variance compared to their contribution in the SSP spectrum. Therefore the main effect of steepening the IMF slope is that we obtain a slightly fainter SBF spectrum. 

It is worth noting that such sensitivities and the fact that elliptical galaxies show radial variations of metallicity, abundance ratios, age and IMF \cite[e.g.][]{MartinNavarro15,MartinNavarro18}, might have a non-negligible impact on the SBF measurements in different regions of a galaxy and resulting SBF gradients (e.g., \citealt{Tonry91,SodemannThomsen95,SodemannThomsen96,LuppinoTonry93}). 
On the other hand, these results point to a potential use of the SBF spectra to constrain relevant stellar population parameters.

Finally we also study the behaviour of the SBF spectra with varying $\aFe$\
abundance. This is illustrated in Fig.~\ref{fig:alpha} for the MILES spectral
range. Overall we find that at constant total metallicity the $\alpha$-enhanced
SBF models are around $0.1-0.2$\,mag brighter in most of the spectral range
covered by the MILES spectral range, being even brighter at the bluest and
reddest ends. On the contrary, the $\alpha$-enhanced model is fainter by
$0.3-0.4$\,mag if the  \Feh\ content is kept constant. In the two cases we find
both a colour term as well as significant variations related to the molecular
bands, throughout the covered spectral range. The obtained differences can be
very useful at aiding SBF distance-based calibrations (see
Section~\ref{sec:distance}).

\subsubsection{Variation with the input isochrones}
\label{subsubsec:evolmod_comp}

In addition to the variation with stellar population parameters, we also test here the effects of using different evolutionary tracks on the SBF spectrum.  Fig.~\ref{fig:sbf_ssp_spec} shows the SBF spectra computed with two sets of isochrones. In particular, the bottom panel shows the ratio of the mean values (SSPs) and the ratio of the SBFs obtained with the Padova00 and BaSTI models. We note that the metallicities between both set of tracks are not identical, however such a difference has a minor impact on this comparison. We find that the ratio of mean values (SSPs) shows almost negligible differences, except in the blue range ($\sim10\%$, percentage of difference). On the contrary the SBF ratio shows significantly larger variations throughout the whole spectral range covered by the models. In fact the SBF ratio highlights differences at short and large wavelength scales, which are related to differences in the molecular bands and in the continuum. These two effects can be attributed in part to differences in the temperature of the RGB evolutionary phase and in the characteristic lifetimes of these stars. We refer the reader to \citet{Vazdekis15} for a more elaborated discussion on the main differences between these two sets of isochrones for the mean SSPs. See also \cite{MFA06} for the effects of varying the isochrones on the SBF magnitudes. The impact of varying the stellar evolutionary prescriptions on the SBF magnitudes is also discussed in \citet{Raietal05,GLL10,GLL18}. 

The observed differences in the SBF spectra translate to differences in the SBF photometry, as shown in Sections \ref{sec:photometric} and \ref{sec:distance}. However the SBF spectra provide us with the advantage of studying these differences with great detail, including how spectral indices would fluctuate according to the selected isochrone set, hence providing an additional test for validating these models. It is worth recalling here that the SBF is a global property of the population, rather than being associated to any given stellar type. Therefore the SBF is more sensitive than the mean value (i.e., the SSP) to the lifetimes of the different stellar types in the ensemble as mentioned above. 

\section{Use of the SBF spectra}
\label{sec:use}

 A relevant peculiarity of the SBF spectrum is that it is not a spectrum in the standard meaning, but rather a ratio between a variance and a mean spectrum. Consequently, these two spectral components must be treated separately in any step, as explained in the following sections. We focus here on some of the possible uses that we envisage for these new models. 
  
\subsection{Star Formation Histories}
\label{sec:SFHs}

We focus here on SBF models that are contributed by different stellar population components. Due to its definition, the SBF spectrum of a composite stellar population cannot be computed by directly combining the single-age single-metallicity SBF spectrum of each component. Instead the mean value and the variance of each SSP must be computed separately and combined afterwards to obtain the resulting SBF spectrum, as their ratio.
  
Prior to the computation of the SBF spectrum of a composite population we first need to {\it i)} combine the mean (SSP) spectra and {\it ii)} calculate the variance of the sum as the sum of the variances, since the different SSPs are not correlated. This is the case because a SSP at a given age is independent from any other SSP with a different age. We therefore proceed to calculate the variance as follows
  
  \begin{equation}
      L^{\mathrm{SFHvar}}_{\lambda}(t) = \int_{0}^{t} \psi(t') \, L^{\mathrm{SSPvar}}_{\lambda}(t'-t)\;\mathrm{d}t' = \sum a_{SFH}(t') \,\sigma^2(t'-t)
      \label{eq:SBF_SFH}
  \end{equation}
  
  \noindent where $\psi(t)$ is the SFH and $a_{SFH}(t)$ are the SFH coefficients when it is discretised as a combination of SSPs. We note that neither $\psi(t)$ nor $a_{SFH}(t)$ contain a squared term. In other words, this is a consequence of the scaling relations of the variance of the stellar population as a function of the stars and mass in the system \cite[c.f.][ and Cervi\~no in preparation]{CLCL06}\footnote{In a back of the envelope argument, the use of $\psi(t)^2$ or $a_{SFH}(i)^2$ coefficients imply that the resulting SBF spectrum would depend on the SFH normalization; whereas using $\psi(t)$ or $a_{SFH}(i)$ the SFH normalization in the variance cancels out with the corresponding SFH normalization of the mean (when is computed the SBF). Another point is that each normalized SSP model provides the probability distribution of a single star, so that by multiplying by $a_{SFH}(i)$ we obtain the total number of stars present in each SSP. Basically, this translates the sum of individual stars into the sum of stars with a given age}. In summary, unlike the SBF spectra, the variance spectra can be treated in the same way as the mean spectra for the computation of composite stellar populations. Finally, it is worth noting that such an approach is valid as far as the stellar birth-rate function can be described by the SFH and IMF, and both functions can be treated separately (i.e. the IMF is independent of the SFH). Note that this is a rather common assumption (see the basic modelling approach described in \citealt{Tinsley80}). If these two functions are not independent of each other it would then become necessary to obtain the correlation among the various SSPs under a given set of stellar birth-rate conditions.

\subsection{Wavelength resampling and spectroscopic SBF magnitudes}
\label{sec:photometric}

   \begin{figure*}
    \includegraphics[width=0.99\textwidth]{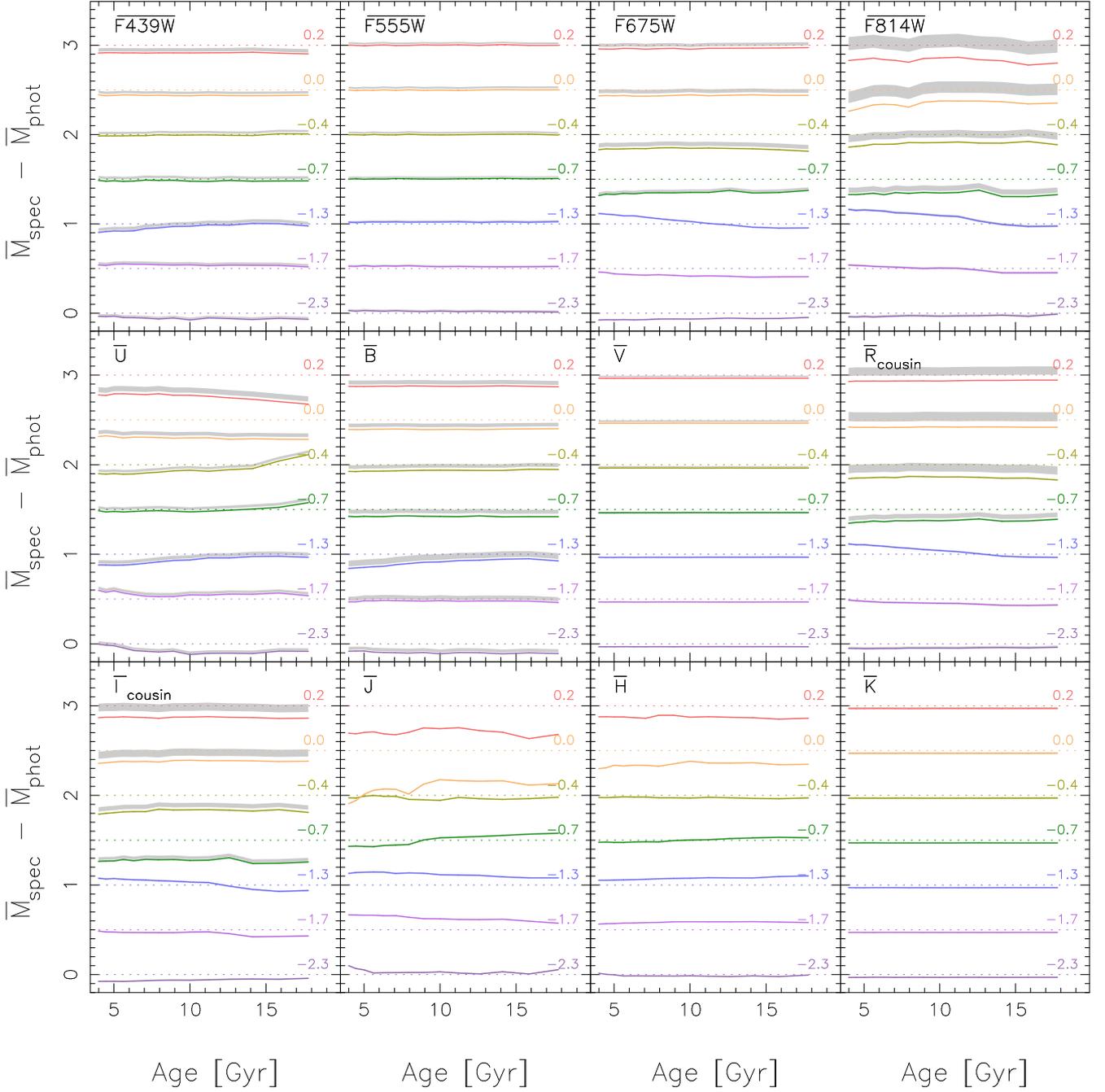}
    \caption{Internal consistency test showing the difference obtained when
comparing the photometric SBF magnitude, Vega system, with the spectroscopic SBF
magnitude for varying filters. The photometric values are obtained on the basis
of extensive photometric stellar libraries, whereas the spectroscopic magnitudes
are obtained by convolving the filter response to the variance and mean
(SSP) spectra, as described in Eq.~\ref{eq:SBF_band2}. For the latter 
the hypothesis of full-correlation between the wavelengths within the filter
passband is adopted. The panels show the results for the various Johnson-Cousin broad-band
filters, except for the top panels, which show the HST WFPC2 filters. The solid
lines show this difference as a function of the stellar population age (in Gyr)
with Padova00 isochrones and Kroupa Universal IMF. The metallicity of these
models, colour coded, increase in each panel from bottom to top, shifted $0.5$
magnitudes for visibility. The metallicity value is shown at the oldest age end
of the perfect agreement black dotted line. The difference obtained after
applying the minimum and maximum offsets due to non full correlation, as
obtained in Appendix~\ref{sec:filter_corr}, are shown by the dashed area.  Note
that for some filters such corrections are so small that it cannot be
distinguished in the plot. Every tick in the vertical axis represents a
fluctuation difference of $0.1\,$mag.}
    \label{fig:phot}%
   \end{figure*}
  
The manipulation of a single SBF spectrum for wavelength re-sampling and other operations concerning wavelengths also differs from the standard treatment of a mean spectrum. This happens when the fluxes in different wavelengths of a single SBF spectrum are combined to obtain a new SBF-related quantity. This applies to the computation of magnitudes from the SBF spectra (hereafter spectroscopic SBF magnitudes), equivalent widths or other spectroscopic indices involving a ratio, or when matching kinematic galaxy properties. In such cases, in addition to the general rule that mean and variance components have to be manipulated separately, it is also necessary to take into account the correlations between different wavelengths.

The main aspect to take into consideration is that we cannot assume that
the SSP variance-flux at a given wavelength $\lambda_i$,
$F^\mathrm{var}_{\lambda_i}$, is independent of the SSP variance-flux at
another $\lambda_j$, $F^\mathrm{var}_{\lambda_j}$. In fact, the difference in
$F^\mathrm{var}_{\lambda}$ for small $\Delta\lambda$ intervals should be
strongly correlated, particularly in adjacent wavelengths, as the contribution
to this flux originates from the same stellar types. Therefore, for a given
integration function described by $T(\lambda)$, comprising a set of
$T_{\lambda_i}$ values with $N$ elements, and using the standard linear
model for variance propagation, we have\footnote{We note that the
trapezoidal rule employed for the integration introduces in the variance an
extra covariance between adjacent bins (see, e.g., Appendix in
\citealt{CastroAlmazan16}). We do not include this effect in
Eq.~\ref{eq:SBF_band0}.}

  \begin{eqnarray}
  \label{eq:SBF_band0}
  F^\mathrm{var}_{band} &=& \sum_{i=1}^N T^2_{\lambda_i} \sigma^2_{\lambda_i} + 2 \sum_{i=1}^N \sum_{j > i}^N T_{\lambda_i} T_{\lambda_j} \sigma_{\lambda_i} \sigma_{\lambda_j}\,\rho_{\lambda_i\lambda_j}\\ \nonumber
  \end{eqnarray}

  \noindent where $\rho_{\lambda_i\lambda_j}$ is the correlation coefficient between the two wavelengths and $\sigma_{\lambda_i} = \sqrt{F^\mathrm{var}_{\lambda_i}}$.  

 As it is rather difficult task to obtain the full set of 
 $\rho_{\lambda_i\lambda_j}$ for each SSP variance spectrum we have explored
 different assumptions, using simple hypotheses about
 $\rho_{\lambda_i\lambda_j}$  in Appendix~\ref{sec:filter_corr}. We used for
 this purpose the {\it SED@} models \cite[][and Cervi\~no in
 preparation]{CLCL06}. We feed these models with the same set of isochrones employed by E-MILES (Padova00 and BaSTI), but with the low resolution BaSeL3.1
 semi-empirical stellar spectral libraries \citep{basel3}. Overall, the best
 results are obtained when assuming the hypothesis of a full correlation between
 wavelengths (i.e. $\rho_{\lambda_i\lambda_j} = 1\, \forall i,j$), which
 transforms Eq.~\ref{eq:SBF_band0} to

  \begin{eqnarray}
  \label{eq:SBF_band}
  F^\mathrm{var}_{band} &=& \left( \sum_{i=1}^N T_{\lambda_i} \sigma_{\lambda_i}\right)^2 \\ \nonumber
      &=&\left( \int T(\lambda)\, \sqrt{F^\mathrm{var}_\lambda}\mathrm{d}\lambda  \right)^2 
  \end{eqnarray}
  
\noindent Finally, we can calculate the spectroscopic SBF magnitude as follows
  
  \begin{equation}
  \label{eq:SBF_band2}
  \bar{m}(\mathrm{spec}) = m_\mathrm{sbf}(\mathrm{spec}) = 2 \times m_{\sqrt{F^\mathrm{var}_\lambda}} - m_{F^\mathrm{mean}_\lambda}
\end{equation}
  
  \noindent where $\bar{m}$ is the standard notation of the SBF magnitude,
  and $m_{\sqrt{F^\mathrm{var}_\lambda}}$ and $m_{F^\mathrm{mean}_\lambda}$ are
  the spectroscopic magnitude of the square root of the SSP variance and the
  mean SSP, respectively. 

We note that in the case of fully correlated wavelengths we obtain the maximum
possible variance. This means that the spectroscopic SBF magnitudes derived in
this way are brighter than the photometric ones. Brighter fluctuations are
unphysical, whereas fainter magnitudes are possible due to non full
correlation within the spectral domain of the filter. Tables
\ref{tab:filter_maxmin_jhonson}, \ref{tab:filter_maxmin_sdss} and
\ref{tab:filter_maxmin_hst} in Appendix~\ref{sec:filter_corr} show the minimum
and maximum offsets obtained, due to non full correlation, for the
Johnson-Cousin, SDSS and HST broad band filters for different metallicities and
ages. These limiting values are provided for each metallicity bin. These values
are obtained by varying the age of the stellar population, for ages above
$4$\,Gyr, and varying the input set of isochrones feeding the models. 
In general, the maximum differences between the spectroscopic and
photometric SBF magnitudes do not exceeded $0.1$\,mag (in most cases they are
lower than $0.04$\,mag), with the exception of the $I$ and F814W/HST bands at
solar and super-solar metallicities, see Figure~\ref{fig:appendix} in
appendix~\ref{sec:filter_corr} for more details. We also refer the reader to 
\cite{CVG03,GLB04} and \cite{SGetal19} for studies about
the correlation between different wavelengths. By simple mathematical
inspection, our approximation should work better for box-shaped filters, which
only affects the integration limits, and fully correlated wavelengths within the
filter domain. It is also more accurate in continuum dominated spectral regions,
or when the spectrum can be approximately represented by very low order
polynomials that, in practise, split the integrated area within the filter
window in two similar parts.

We made use of Eq.~\ref{eq:SBF_band2} to obtain the spectroscopic SBF
magnitudes. To perform an internal consistency model check, we compared 
these fluctuation magnitudes to those computed with the same stellar population
synthesis code but following the classical approach, which integrates the fluxes
in these bands \citet{Blakeslee01}, i.e. hereafter the photometric E-MILES
predictions. For computing the E-MILES photometric fluctuation magnitudes
we employ extensive photometric libraries, rather than spectra. These libraries
are the same ones that we use to transform the theoretical parameters of the
isochrones to the observational plane, as described in detail in
Section~\ref{sec:isochrones}. Figure~\ref{fig:phot} shows the resulting
fluctuation magnitude difference as a function of age and metallicity for
representative filters from the Johnson-Cousin and HST WFPC2 Vega system, as
published in \citet{Blakeslee01} and updated here. In each panel the metallicity
of the models is colour coded and shifted by $0.5$ mag for visibility. The solid
line is the difference between the spectroscopic fluctuation magnitudes, using
Eq.~\ref{eq:SBF_band2}, with respect to the corresponding photometric
fluctuation magnitudes. The dotted line represents a perfect match, whereas
the dashed area corresponds to the spectroscopic values after applying the
minimum and maximum offsets as obtained in Appendix~\ref{sec:filter_corr}, due
to non full correlation between the wavelengths within the band. Such a
comparison also helps in assessing the relative impact of the varying empirical
libraries that feed our models for both, the photometric and the spectroscopic
predictions. Besides the full correlation approximation adopted here for
obtaining the spectroscopic SBF magnitudes, i.e. Eq.~\ref{eq:SBF_band2}, we do
not expect a perfect match. This is because the photometric and spectroscopic
libraries differ in their stellar parameters coverage and other aspects such as,
e.g., flux-calibration issues affecting the stellar spectra. Such aspects are
likely responsible for the additional residuals that contribute to the mismatch
seen in some filters. These additional offsets are larger than the ones due to
non full correlation effects for the $F814W$ and $R$ filters, similar for the
$F439W$, $F675W$, $U$, $B$ and $I$ filters, and negligible for the remaining
ones. Note, however, that these offsets are calculated with the aid of the {\it
SED@} models, which make use of a semi-theoretical library. Indeed, the coverage
of this semi-theoretical library for cool stars is scarce in comparison to that
in E-MILES. Such issues might introduce further residual fluctuation
differences, which prevent us from separating more cleanly these two major sources
of uncertainties. Moreover, non full covariance effects seem to be
underestimated in several filters, and particularly in the $J$ and $H$ bands.
For comparison, we note here that our photometric and spectroscopic magnitudes
for the mean SSPs lead to differences that are typically within $0.02$\,mag
\citep{MIUSCATII}.

Overall, the fluctuation magnitudes agree within $0.1\,mag$, and are smaller
in most cases, but can also be as large as $0.2\,mag$ for some filters. The
spectroscopic SBF magnitudes obtained for the HST filter system show in general
the best match to the photometric values. This is not surprising given the fact
that these filters have a significantly more boxy-shaped transmission. The
very best agreement is reached for the $V$ and $K$ bands for all metallicities
and ages. This is not surprising given the procedure adopted to scale the
individual stellar spectra during the computation of the SSP, as described in
Section \ref{sec:SBF_spectra}. The largest differences are found for the $J$
band filter, particularly in the high metallicity regime. These fluctuation
magnitude differences can be attributed in part to the huge molecular bands 
that are present in the $J$ band of the empirical stars that feed E-MILES. 
Such features are not reproduced by the low resolution theoretical atmosphere
models that feed the {\it SED@} models, leading to inaccurate offsets. We also
find significant differences in the $U$ and $H$ filters, primarily for the most
metal-rich stellar populations. The differences obtained for the $H$ band can be
attributed to the same reasons affecting the $J$ band. However, the differences
affecting the metal-rich spectroscopic SBF magnitudes obtained for the $U$
band not only originate in the strong and wide spectral features present in
this band but, also, in the very steep spectrum shape that characterise the SBF
spectra of old stellar populations. We discourage potential users of these SBF
spectra to obtain spectroscopic magnitudes in the UV spectral region. 
Despite these limitations we stress here that the SBF spectra provide the
flexibility that is required to convert SBF magnitudes obtained with varying
photometric systems. This is particularly useful for homogenizing data coming
from different sources. In fact, by applying Eq.~\ref{eq:SBF_band2} for a given filter definition to two different observational setups one can use the obtained spectroscopic magnitude difference to bring one photometric measurement to the other. Of course, this differential correction is more accurate when using the SBF spectra that correspond to the best matching mean values.

\section{Comparison to other SBF spectral models}
\label{sec:others}

We have compared our model spectra to the recently published SBF model spectra of \citet{Mitzkus18}. A unique and common feature of this comparison is that these authors also employ the MILES stellar database. For the other major ingredient of their population synthesis code, these authors make use of the PARSEC isochrones \citet{Bressan12}. These isochrones are a revised version of the former Padova models, with updated input physics and using the solar composition from \citet{GrevesseSauval98} and elemental abundances from \citet{Caffau11}. For the TP-AGB stars these models make use of the COLIBRI code \citep{Marigo17}. Finally, it is worth noting that although these authors also compute SBF spectra based on a theoretical stellar library, namely that of \citet{Coelho14}, we use here their version based on MILES to minimize the differences with our models. Their models are computed with the \citet{Chabrier03} IMF. 

\begin{figure}
    \includegraphics[width=0.49\textwidth]{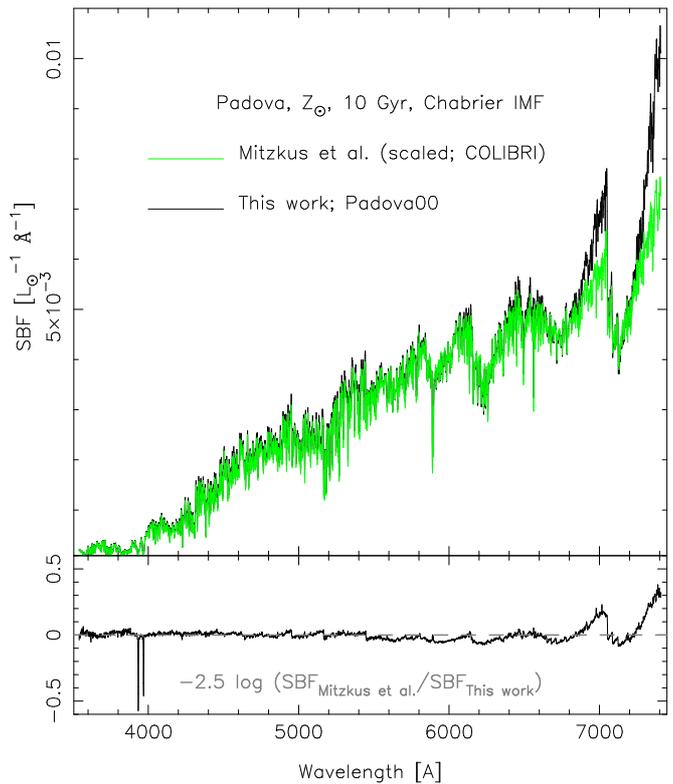}
    \caption{Comparison of our model SBF spectrum (black) with solar metallicity, $10$\,Gyr and Chabrier IMF, with the SBF spectrum of \citet{Mitzkus18} (green) with similar stellar population parameters. In the bottom panel we show the residuals.}
    \label{fig:comp_Mit}%
   \end{figure}

Figure~\ref{fig:comp_Mit} shows a representative SBF spectrum from these authors with solar metallicity and $10$\,Gyr, compared to an equivalent model computed with our code with the Padova00 isochrones. As the \citet{Mitzkus18} spectra are not calibrated in flux, but the relative spectral shape, we scaled their SBF spectrum by an arbitrary factor to match our models. Overall we find a good agreement, although there are significant residuals at intermediate wavelength scales redward of $\sim6000$\,\AA. These residuals are associated with the molecular bands that are present in evolved cool stars. According to Fig.~12 in \citet{Bressan12} the brightest and coolest RGB stars in the PARSEC models are cooler than in the Padova00 that are employed in our models, although the slope of this evolutionary phase is steeper in Padova00, which makes the base of the RGB slightly cooler. These differences might suggest that the SBF spectrum by \citet{Mitzkus18} presents deeper absorption molecular bands compared to ours,  as the SBF goes as the inverse of the mean SSP spectrum. There are also other differences between the two stellar population synthesis codes, such as, e.g., the algorithm employed to assign stars from the MILES database when integrating along the isochrone. The fact that we do not see any significant colour term in the obtained residuals, but differences in the molecular bands in the red spectral range, strongly suggest that differences in the SBF spectra mostly originate in the modelling approaches related to the coolest stars near the tip of the RGB.

  \begin{figure}
    \includegraphics[width=0.49\textwidth]{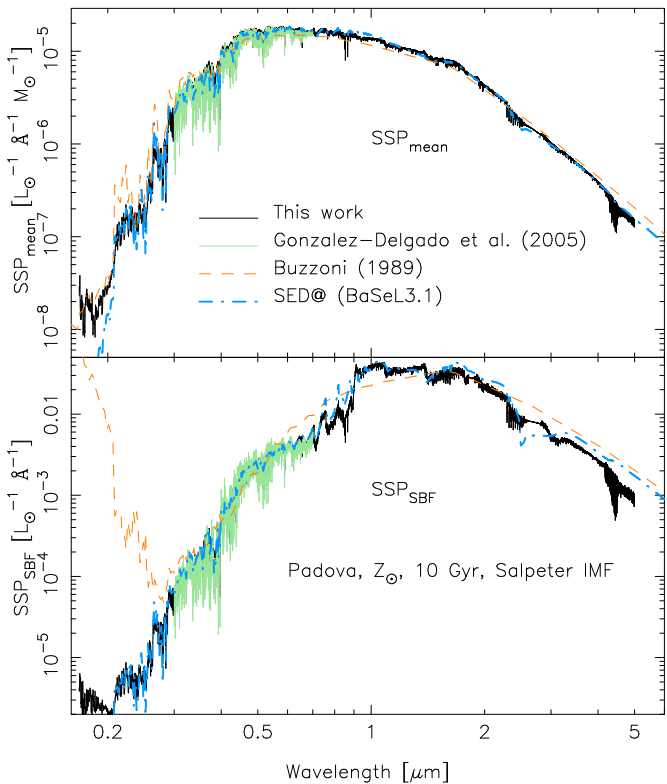}
    \caption{Comparison of our models (solid black) with solar metallicity, $10$\,Gyr and Salpeter IMF, with the models of \citet{Buzz89} (dashed orange) and \citet{GDetal05} (solid green) with similar stellar population parameters. We also include in this comparison the low-resolution {\it SED@} models (dot-dashed light blue) employed in Section~\ref{sec:photometric}.}
    \label{fig:comp_other}%
   \end{figure}

We also compared our SBF spectra with those of \citet{Buzz89}\footnote{{\tt http://www.bo.astro.it/$\sim$eps/models.html}} and the {\it SED@} models presented in \citet{GDetal05}\footnote{{\tt https://www.iaa.csic.es/$\sim$rosa/research/synthesis/HRES/ESPS-HRES.html}} \footnote{\tt http://cab.inta-csic.es/users/mcs/SED/}. Both set of models provide the SSP variance, although encoded as the effective number of stars ${\cal N}_\lambda$. The relation between $\cal N$, SBF and mean SSP spectra is given by \cite[c.f.][]{Buzz93}:

\begin{equation}
    L_{\lambda}^\mathrm{SSPmean} = L_{\lambda}^\mathrm{SSPvar} \times {\cal N}_\lambda
    \label{eq:Neff}
\end{equation}

\noindent All these models are computed with the Salpeter IMF (in our case the
unimodal IMF option with logarithmic slope 1.35) and are calibrated in flux, so
that we can directly compare the absolute values. However, the stellar spectral
libraries (theoretical in the case of these authors) and the isochrones differ
between the models. The \citet{Buzz89} models include the evolution of Post-AGB
stars, three different morphologies for the Horizontal-Branch and two different
mass-loss parameters $\eta$. For the comparison with our models we selected the
models with red HB morphology and $\eta=0.3$. The {\it SED@} models presented in
\citet{GDetal05} make use of the theoretical stellar spectral library of
\citet{Maretal05} and the Padova00 isochrones. Figure~\ref{fig:comp_other}
shows the SBF spectra from these authors, together with the E-MILES models. All
these models were computed with the Padova00 isochrones with Salpeter IMF,
$10$\,Gyr and solar metallicity. In addition, we also plot the low-resolution
{\it SED@} model employed here, which was used to estimate the effects of the
full correlated spectrum approach described in Section~\ref{sec:photometric}. In
general, the models agree reasonably well among each other, given the varying
model ingredients \cite[we refer to][~for a more detailed study about the
variations on stellar libraries]{Mitzkus18}.

\section{Applications}
\label{sec:applications}

We show here some applications of the newly computed SBF spectra. Specifically we focus on the use of SBF magnitudes derived from these spectra.
 
\subsection{Comparison with empirical SBF distance calibrations} 
\label{sec:distance}

The determination of galaxy distances has been a central problem in Astronomy.
One of the most accurate methods to obtain distances is using SBFs as the
observed flux of the fluctuations depends on the distance of the object, with
more distant galaxies appearing smoother. The SBF method has given distances to
Virgo and Fornax with a precision of $2\%$ \citep[see][for a
review]{Blakeslee12}.

\begin{figure}
    \includegraphics[width=0.49\textwidth]{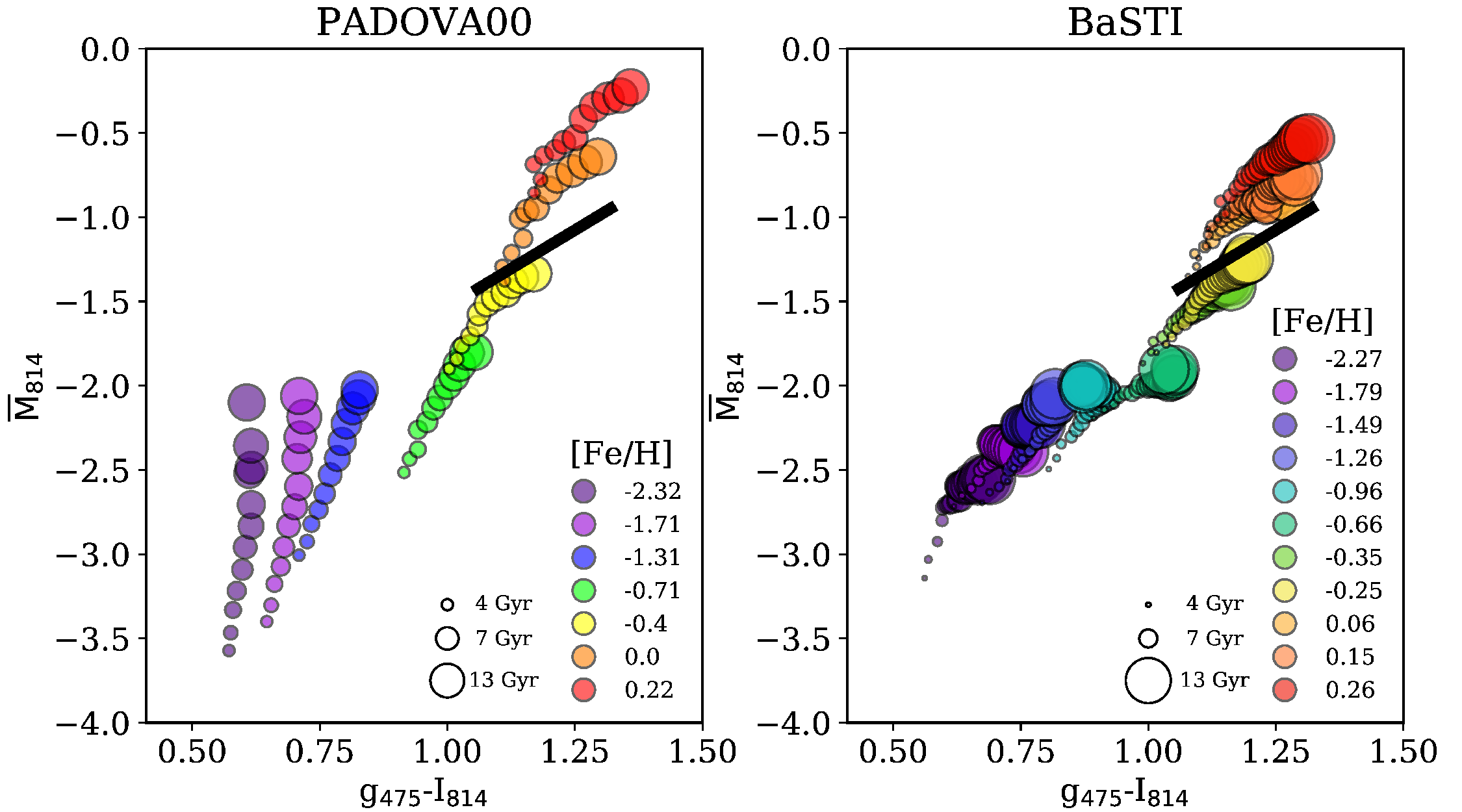}
    \caption{Absolute fluctuation (AB) magnitude in the $F814W$ ACS/HST filter vs. the $g_{475}-I_{814}$ colour. The thick black line shows the calibration of Blakeslee et al. (2010) (their Eq.~2) within its validity range. The filled circles show our model predictions based on the Padova00 (left panel) and BaSTI (right panel) stellar models, computed with bimodal IMF with slope 1.3. The metallicity is varied according to the colour-coding quoted in the bottom right side of the panel, while the sizes of the circles increase with age.}
    \label{fig:Blakeslee_calibration}%
   \end{figure}

To measure  distances, observational studies have aimed at accurately estimating
the absolute SBF magnitude, $\bar{M}$, using empirical calibrations. However, as
$\bar{M}$ is an intrinsic property of the stellar populations of a galaxy, it is
possible to derive these calibrations from stellar population models. 
Figure~\ref{fig:Blakeslee_calibration} shows the absolute SBF magnitudes in the
F814W/ACS/HST band, $\bar{M}_{814}$, derived from SBF spectra as a function of
$g_{475}-I_{814}$ colour. $\bar{M}_{814}$ is obtained following the methodology
described in Sec.~\ref{sec:photometric} using the response of the F814W/ACS/HST
filter.  The mean spectra corresponding to these SSPs were used to obtain the
spectroscopic $g_{475}$ and $I_{814}$ magnitudes by direct 
convolutions with the F475W and F814W/ACS/HST filters, respectively. The
models used here and shown in Fig.~\ref{fig:Blakeslee_calibration} adopt a
bimodal IMF with a slope of 1.3 for the Padova00 (left panel) and BaSTI (right
panel) isochrones. Each point is colour-coded according to its metallicity from
purple, \Mh$=-2.3$, to red, \Mh$=0.26$. Symbol size increases with age, and we
plot ages from $4$ to $13$ Gyr. The solid black line is the calibration given in
Eq.~2 in \citet{Blakeslee10}, derived from early-type galaxies in the Fornax
cluster. Contrary to what is expected for massive ETGs, this calibration lies at
metallicities slightly lower than solar metallicity. 

Apparently these lower inferred metallicities might be an indication of
metallicity gradients, as observed in elliptical galaxies
\citep[e.g.,][]{Montes14,MartinNavarro18}, which lowers the total integrated
metallicity observed in these galaxies. \citet{Blakeslee10} measured their SBF
magnitudes at radii between $6\farcs4$ to $32$\arcsec~where steep metallicity
gradients are observed in early-type galaxies
\cite[e.g.,][]{SanchezBlazquez07,Spolaor10}. In fact,
\citet{Cantiello05,Cantiello07} found that the size of the internal SBF-colour
slope for multiple annuli in their sample of galaxies observed with ACS/HST is
related in most cases to  metallicity gradients, although in some cases to
age gradients.  Since the presence (or absence) of a SBF gradient is connected
to the properties of the dominant stellar population, it could be used as a
tracer of the formation scenario of the galaxy \citep{Cantiello11}. A more
detailed discussion on the connection between these gradients and the derived
SBF magnitudes can be also found in, e.g., \citet{Tonry91,Jensen98,Cantiello05}.
Such gradients should be properly taken into account when deriving SBF
magnitudes in galaxies for very precise distance determinations. The SBF signal
and the galaxy colour should be obtained in the same region of the galaxy.

The fact that our SBF models point toward lower metallicities can be attributed
in part to the physical ingredients in stellar evolution, and stellar libraries,
employed for computing the SBF spectra, from which these magnitudes are derived.
This is shown in Fig.~\ref{fig:Blakeslee_calibration}, where the fluctuation
magnitudes predicted on the basis of the Padova00 models place the empirical
distance calibration at relatively lower metallicities than those from the BaSTI
models. Moreover, the fact that the most massive galaxies show \MgFe-enhancement
(e.g., \citealt{Renzini06}) might have non-negligible implications. In
Fig.~\ref{fig:alpha} we showed that at constant total metallicity the
\MgFe-enhanced models with solar metallicity are brighter than their
solar-scaled counterparts by at least $0.2$\,mag around $\sim7000$\,\AA.
Correcting this difference will bring our solar metallicity models on the top of
the distance SBF calibration. These results are along the lines of those shown
by \citealt{Lee10}, who found that their $\alpha$-enhanced models are
$0.35$\,mag brighter than the solar-scaled models in the $I$ band. 
A caveat has to be taken into account at this point as
the $R$ and $I$ band spectral regions, and in general the wavelength interval
around $7000$\,\AA, are heavily affected by correlation effects. In fact,
the $F814W$ band shows the largest differences between spectroscopic and
photometric SBF magnitudes at the solar and super-solar metallicity regimes, as
clearly shown in Fig.~\ref{fig:appendix} and Fig.~\ref{fig:phot}. Therefore the
expected brightening of the SBF due to the consideration of \alfa-elements
abundances might not be sufficient to be able to match the empirical
calibrations.

\begin{figure}
    \includegraphics[width=0.49\textwidth]{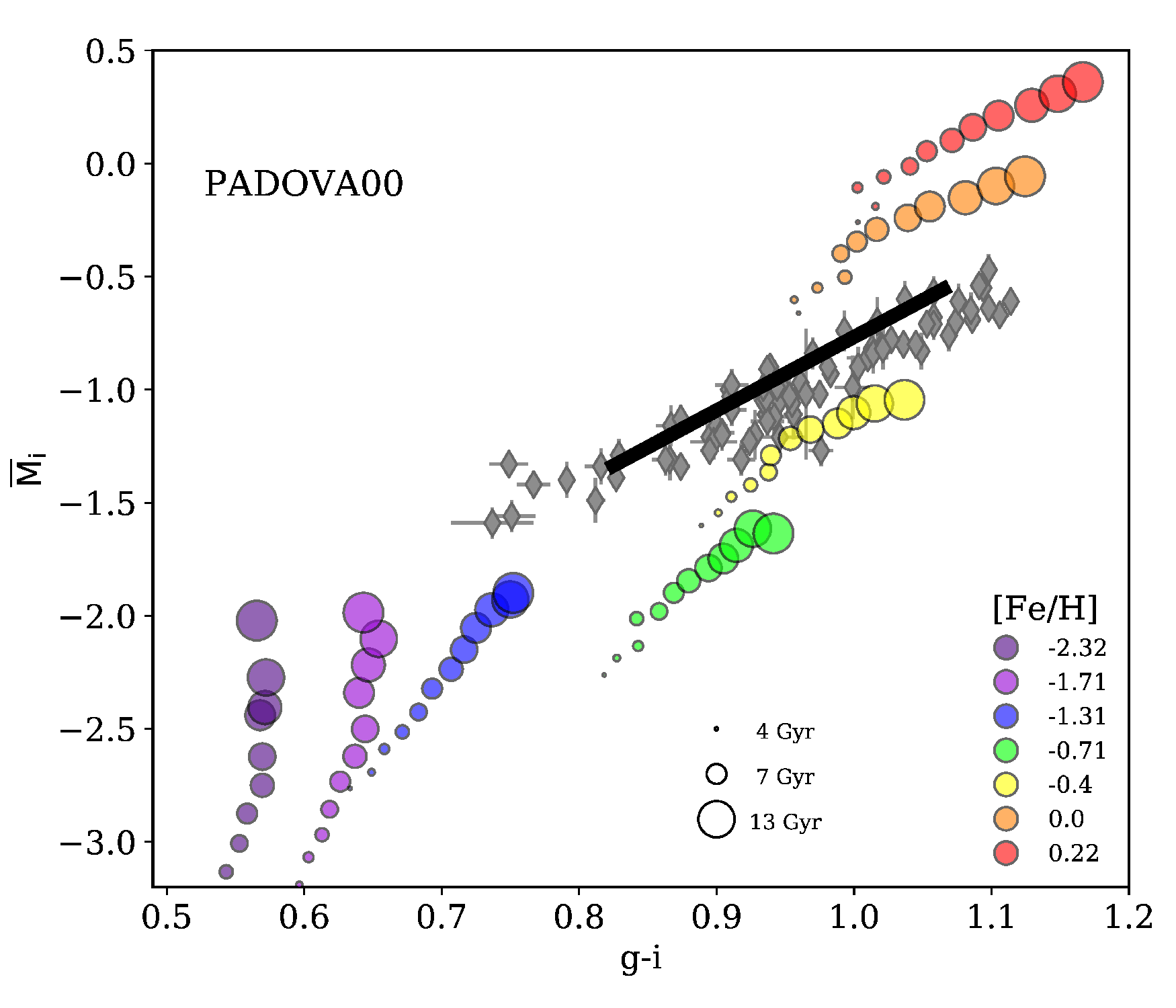}
    \caption{Absolute $i$ band fluctuation as a function of the $g - i$ colour. Both galaxy points (shown in grey) and the calibration within its validity range (thick black line) are taken from \citet{Cantiello18}. The filled circles show our model predictions adopting a bimodal IMF with slope 1.3 and Padova00 stellar models. The metallicity is varied according to the colour-coding quoted in the bottom right side of the panel, while the sizes of the circles increase with age.}

    \label{fig:Cantiello_calibration_Padova00}%
   \end{figure}

\begin{figure}
    \includegraphics[width=0.49\textwidth]{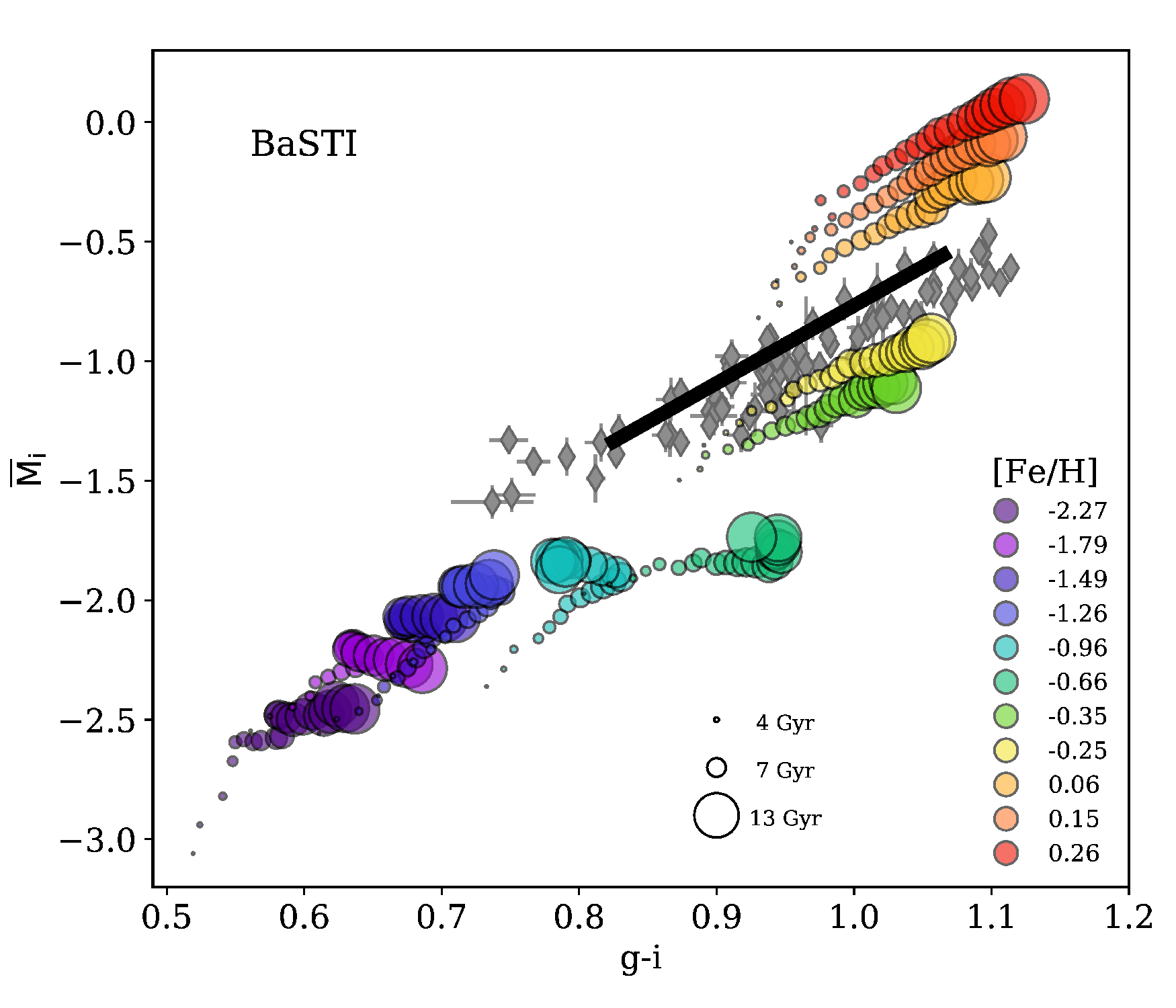}
    \caption{Same as Fig.~\ref{fig:Cantiello_calibration_Padova00} but for the models computed with the BaSTI isochrones.}
    \label{fig:Cantiello_calibration_Teramo}%
   \end{figure}

Recently, \citet{Cantiello18} used MegaCam/CFHT imaging of the Virgo cluster to
measure SBF distances and provided multiple calibrations of the SBF absolute
magnitude in the $i$-band for different colour combinations. In
Fig.~\ref{fig:Cantiello_calibration_Padova00} and
Fig.~\ref{fig:Cantiello_calibration_Teramo}, we plot the absolute SBF magnitudes
in the MegaCam/CFHT $i$-band as a function of the $g-i$ colour, for the same
models described above. The calibration (black line)\footnote{The empirical
calibration derived in \citet{Cantiello18} is derived using a subsample of the
data.} and data (grey diamonds) for galaxies in the Virgo cluster taken from
\citet{Cantiello18} are also plotted. Note that the calibration in
\citet{Cantiello18} extends to bluer colours than the \citet{Blakeslee10} one.
Regarding the behaviour of SBF magnitudes at lower metallicities,
\citet{Blakeslee09} found that the scatter in their empirical calibration
increased at the bluest colours \citep[see also][]{Cantiello18}. The fluctuation
magnitude measurements in a sample of $25$ dwarf spheroidal galaxies in the
Fornax cluster obtained by \citet{Mieske06} yielded a cosmic scatter of
$0.34$\,mag. This scatter is significantly larger than that found at redder
colours. This is consistent with what we find for our models at [Fe/H] $< -1$,
where the values of the SBF absolute magnitude are not uniquely defined with
respect to their integrated colours. This scatter might be driven by the
presence of stellar populations with different metallicities and ages (in
\citealt{Cantiello18} most of the galaxies at $g-i<0.9$ are late-type). See also
the discussion in \citet{Trujillo19} for the case of an ultra diffuse galaxy and
\citet{Carlsten19} for satellite galaxy systems. Indeed, it has already been
suggested that in order to explore SBF calibrations at bluer colours it is
necessary to use nonlinear calibrations involving multiple colours
\citep[e.g.][]{Blakeslee09, Cantiello18}.

Our spectroscopic SBF magnitudes are able to match fairly well the empirical SBF distance calibrations, within their validity range. The advantage of using theoretical SBF calibrations is that they are less time consuming to derive, homogeneous in different bands and they make SBF distances a primary distance indicator as they do not rely on Cepheids distances as a zero point. We remark here that deriving SBF based distances with precision require more accurate procedures that the tests performed here. However our SBF spectra are promising and have a potential in aiding these studies based on very precise photometric fluctuation magnitudes.
 
Finally, we would like to emphasize the utility of the SBF spectra synthesized here to transform between different photometric filter definitions and systems employed for the distance measurements shown here. Moreover, with these models it is possible to achieve more precise comparisons with fluctuation magnitudes and colours measured at varying recession velocities.

\subsection{Disentangling relevant stellar population parameters: metal-poor contributions} 
\label{sec:metal-poor}

   \begin{figure}
    \includegraphics[width=0.49\textwidth]{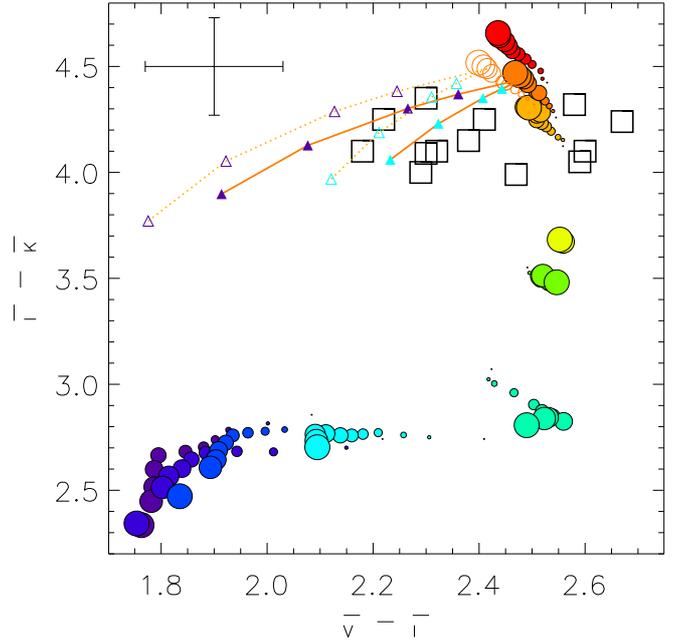}
    \caption{The $I - K$ fluctuation colour is plotted against $V - I$ fluctuation colour (Vega system) for a set of representative ETGs (black open squares with the mean error bars shown on the top left corner; see the text for the sources). Single-burst models (filled circles) of increasing size, which represent an age increase from $4$ to $14$\,Gyr, are shown for each metallicity, \Mh$=-1.79, -1.49, -1.26, -0.96, -0.66, -0.35, -0.25, +0.06, +0.15$ and $+0.26$, from blue-purple to red, respectively, which are easily distinguished by the various model loci. We adopt a standard "bimodal" IMF with logarithmic slope $1.3$ for the upper-segment (above $0.6$\,\Msol). For models with \Mh$=0.15$ we also show the predictions for a  bottom-heavy bimodal IMF with slope 2.8 (open orange circles). In addition, we show composite populations made with a dominant metal-rich component, \Mh$=0.15$, with increasing mass-fractions ($1$, $2$, $5$ and $10\%$) of a rather metal-poor component (\Mh$=-1.79$ blue-purple filled triangles and \Mh$=-0.96$ in cyan filled triangles). Similar combinations are shown for a bottom-heavy IMF (open triangles with the same colour criterion). These combinations are joined by an orange solid line (standard IMF) or dotted line (bottom-heavy IMF). In all cases the mass-fraction of the metal-poor component increases from redder to bluer $V - I$ fluctuation colour, with these lines starting from the SSP value, i.e. a nil contribution from a metal-poor population.}
    \label{fig:SBFcolorcolor_T}%
   \end{figure}   

Given their connection to the properties of the stellar population, SBFs provide additional information on the formation scenarios of galaxies. Fluctuation magnitudes and colours have been used previously to constrain relevant stellar population parameters (e.g., \citealt{Liu00}). To illustrate a possible application of these new models, we compare our fluctuation colour-colour predictions, as derived from the newly synthesized model spectra, with data of a representative set of early-type galaxies, taken from \citet{Blakeslee01}. Such fluctuation colour-colour diagrams were already employed by other authors (e.g., \citealt{Liu00,Blakeslee01}), but its use have been hindered by the small number of SBF measurements for varying filter bands in the same galaxies. 
The SBF spectra computed here provide a flexible way to obtain spectroscopic magnitudes that match observations obtained with any filter/system. We follow the approach described in Section~\ref{sec:photometric} to obtain the fluctuation colours that match the observational setup.

Figure~\ref{fig:SBFcolorcolor_T} shows the $I - K$ vs. $V - I$ fluctuation colour-colour diagram, with a set of representative ETGs (black open squares). The $V$ galaxy data are from \citet{Tonry90} and \citet{Blakeslee01}, the $I$ data are from \citet{Blakeslee99}; the $K$ data are from \citet{LuppinoTonry93} (two galaxies), \citet{Jensen98} (10 galaxies) and \citet{PahreMould94} (one galaxy). The plotted single-burst SBF predictions cover a range of metallicities ($-1.79<\Mh<0.26$) and ages above $4$\,Gyr. We show the predictions for a standard "bimodal" IMF with logarithmic upper-segment slope $1.3$ (above $0.6$\,\Msol), but we also show some predictions with a bottom-heavy bimodal IMF with slope 2.8 for models with \Mh$=+0.15$ (open orange circles). This choice of IMF is motivated by recent claims suggesting such dwarf-enriched IMFs in the central regions of massive elliptical galaxies (e.g., \citealt{LaBarbera13,MartinNavarro15}). We see that the data fall in a region that is not matched by the models, which seem to loop to around the observational data, with a mean $V - I$ fluctuation value around $\sim2.4$. The single-burst SBF values only approach those galaxies with $V - I$ fluctuation colour above $2.4$ (within the error bars). However M\,32, with the reddest value, seems to require stellar populations with ages below $4$\,Gyr, in good agreement with the detailed analysis of its mean spectrum \citep{Rose05}. Varying the IMF shifts the SSP values by $\sim0.05$\,mag in the correct direction, i.e. along the $V - I$ fluctuation axis as indicated by the open orange circles, with no significant change in the  $I - K$ fluctuation axis. However the galaxies with $V - I$ fluctuation value smaller than $\sim2.3$ remain clearly out of the reach of the model $V - I$ fluctuation values. 

Although not shown we also tested whether the fluctuation colours corresponding to younger ($<4$\,Gyr) SSPs would be able to match the observed galaxy values. We did not find any single model that could reproduce the galaxies with the bluest $V - I$ fluctuation colours, except for a very few models with $\Mh=-0.35$ and $\Mh=-0.25$, which provide $V-I$ fluctuation values that approached a fluctuation colour value of $\sim2.3$. However these contributions would have a large impact on the mean optical galaxy spectrum, such as, e.g., strengthening the Balmer lines, that would not match the observations. A full fitting that comprises varying fractions of components with different ages and metallicities is out of the scope of this work. However we emphasize here the great advantage of using in the analysis both the mean and the SBF spectrum to be able to properly constrain the inferred solutions.

The model loci in Fig.~\ref{fig:SBFcolorcolor_T} strongly favour the presence of some contributions from fairly metal-poor stellar populations. Therefore we investigate the effects of varying mass fraction contributions from a metal-poor component on  top of a dominant metal-rich population. For this purpose we combine an old population of $12$\,Gyr and slightly super-solar metallicity (\Mh$=+0.15$) with varying mass-fractions (from $1$ to $10\%$) of a similarly old component with two different metallicity values $-1.79$ (blue-purple triangles) and $-0.96$ (cyan triangles). The models including the component with $-1.79$ require mass-fractions around $\sim1-2\%$ on the top of a  vastly dominant super-solar stellar population with a bottom-heavy IMF (open triangles), and $\sim2-4\%$ with a standard IMF (filled triangles). If the metallicity of the metal-poor component is \Mh$=-0.96$ we obtain mass-fractions as high as $10\%$ when the IMF is standard.
We also note that the derived mass fractions for the metal-poor component are not significantly different when we adopt solar metallicity for the dominant stellar population. This is not surprising given the fact that the SSP models of this metallicity provide $V-I$ fluctuation colour values that are close to the super-solar ones.

We have shown that this approach, based on the fluctuation colours, is able to disentangle contributions from very metal-poor components. However this analysis is not completely free from degeneracies as there are varying combinations of stellar populations leading to the same results. In summary, we find that the obtained results depend mostly on the adopted metallicity for the metal-poor component, leading to mass-fractions that can be almost a factor $\sim10$ times larger. 
Note that we have assumed simple models composed of two populations, one old metal-rich to represent the overwhelming dominating contribution and the other one for the very metal-poor, also old, component that characterizes the stellar populations formed during a rapid chemical enrichment that took place before reaching the peak of formation \citep{Vazdekis96,Vazdekis97}. Therefore the larger (smaller) mass-fractions derived when adopting less (more) extreme low metallicities is indicative of a distribution in metallicity that represents that initial enrichment stage. Although out of the scope of this paper we envisage that a more realistic model description for this phase will likely provide a more robust cumulative mass fraction for this metal-poor component. We emphasize that these mass fraction estimates must be taken as an upper limit as our models do not incorporate Post-AGB star spectra that, unlike in the mean SSP spectra, raise the flux of the SBF spectra blueward $\sim3000\,\AA$. This is shown in Fig.~\ref{fig:comp_other} for the models of \citet{Buzz89}. However such contribution has virtually a negligible impact on the optical range and particularly for the $V-I$ fluctuation colour employed in Fig.~\ref{fig:SBFcolorcolor_T}.
 
Finally, we also find that when changing the adopted IMF for the dominant component, this fraction varies by a factor of $\sim2$. Finally, the age of the dominant population has little impact on the derived fractions. Additional constraints can be obtained if combined with the analysis of the mean stellar populations properties such as line-strength indices and high resolution spectra.

Our purpose here is to show the potential of the SBF analysis to separate these small contributions, which warrants further research to provide more constrained quantitative estimates. It is worth mentioning that such metal-poor contributions are very difficult to constrain from the standard analysis based on the mean-SSP spectra as a young contribution has a similar effect on the Balmer line indices as the old metal-poor component (e.g., \citealt{MarastonThomas00}). The mass-fractions derived here are in rather good agreement with the full chemo-evolutionary theoretical predictions of \citet{Vazdekis97}. According to these models, such metal-poor fractions come from the early stages of galaxy evolution during a rapid chemical enrichment phase. Note that such models were able to fit a large variety of colours and line-strength indices of massive ETGs under the assumption of a closed box modelling approach, i.e. fully in-situ formation process.

   \begin{figure}
    \includegraphics[width=0.49\textwidth]{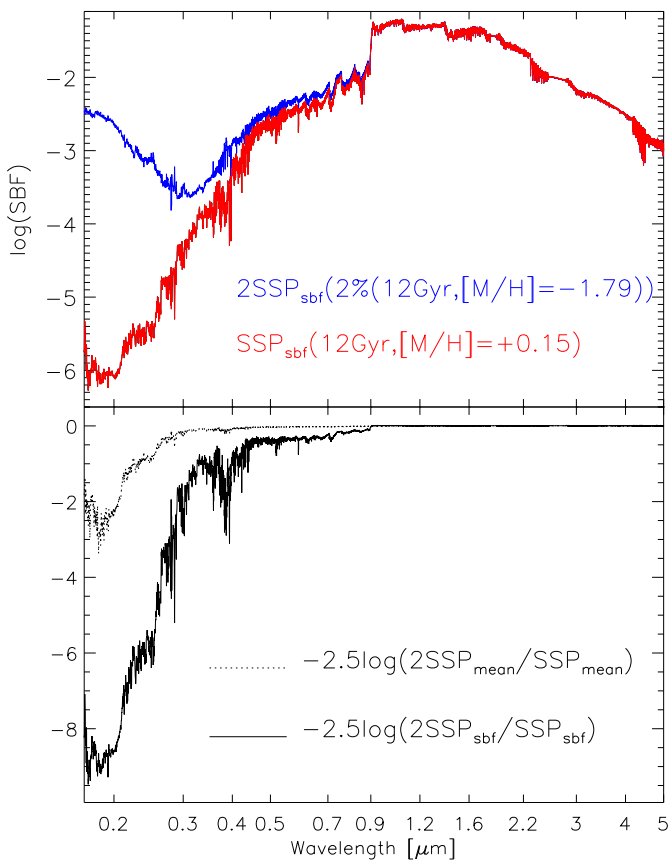}
    \caption{The upper panel shows the SBF spectrum corresponding to an SSP with age $12$\,Gyr and super-solar metallicity \Mh$=+0.15$ (plotted in red). Also shown is a representative solution for the colour-colour fluctuations diagram of Fig.~\ref{fig:SBFcolorcolor_T} (plotted in blue), namely, the composition of that old super-solar SSP with a $2\%$ mass-fraction contribution from a population with metallicity \Mh$=-1.79$ and similar age. The lower panel shows the difference in magnitudes between these two SBF models (solid black line) and their corresponding mean models (dotted black).}
    \label{fig:SBF_poor}%
   \end{figure}   

A significant step further can be achieved by analyzing galaxy SBF spectra, as  has been recently attempted by \citet{Mitzkus18}. In Fig.~\ref{fig:SBF_poor} we illustrate this by showing an SBF spectrum corresponding to a single SSP with old age and super-solar metallicity, as typically found for massive elliptical galaxies through a standard spectral analysis, and the SBF spectrum corresponding to a representative solution in the colour-colour fluctuations diagram shown in Fig.~\ref{fig:SBFcolorcolor_T}. The latter model is composed of an old super-solar SSP with a $2\%$ mass-fraction contribution from a population with metallicity \Mh$=-1.79$ and similar age. The impact of this contribution is barely detected redward the $I$ band, but becomes very strong in the $U$ band, leading to a $\sim1$\,mag difference with respect to the magnitude of the largely dominant old metal-rich population alone. Note that this capability of the SBF spectra to recover such small fractions of very metal-poor contributions is not related to the age-metallicity degeneracy affecting the mean spectra, in its classical understanding. This degeneracy refers to the overwhelming dominant old metal-rich component that can either be more metal-rich and younger or the other way around to be able to match the observed mean colours.

Another interesting result is the significant flattening of the SBF spectrum shape in the $V$ and $R$ bands, which is responsible for the blueing of the $V - I$ fluctuation colour in Fig.~\ref{fig:SBFcolorcolor_T}. In contrast, the impact of this metal-poor contribution is almost negligible in the mean spectra corresponding to the combined population, being smaller than $0.1$\,mag redward $\sim3000$\,\AA. We also see that the impact of the metal-poor contribution in the UV spectral range is extremely significant, leading to a difference that becomes as large as $\sim8$\,mag around $\sim2000$\,\AA. In this spectral range the effect of this metal-poor component in the corresponding mean SSP spectra is also relevant, reaching $\sim2$\,mag. However, it would be very difficult to separate these components on the basis of the mean stellar population spectra, as the NUV is also sensitive to tiny contributions from rather young populations \citep{Vazdekis16,Salvador-Rusinol19}.

Finally, although not shown, we obtain similar residuals to those shown in the lower panel of Fig.~\ref{fig:SBF_poor} for other representative solutions, such as that with a $5\%$ mass-fraction contribution of a population with metallicity \Mh$=-0.96$. In a more realistic case it is expected that galaxies include such small contributions for a range in metallicities with \Mh$<-1$. In this case the net mass fraction corresponding to these populations cannot exceed significantly that of a single population with \Mh$=-0.96$, i.e. $\sim10\%$ as inferred from Fig.~\ref{fig:SBFcolorcolor_T}. 

\subsection{Narrow filter SBF SEDs} 
\label{sec:Narrow_filters}

   \begin{figure*}
    \includegraphics[width=0.9\textwidth]{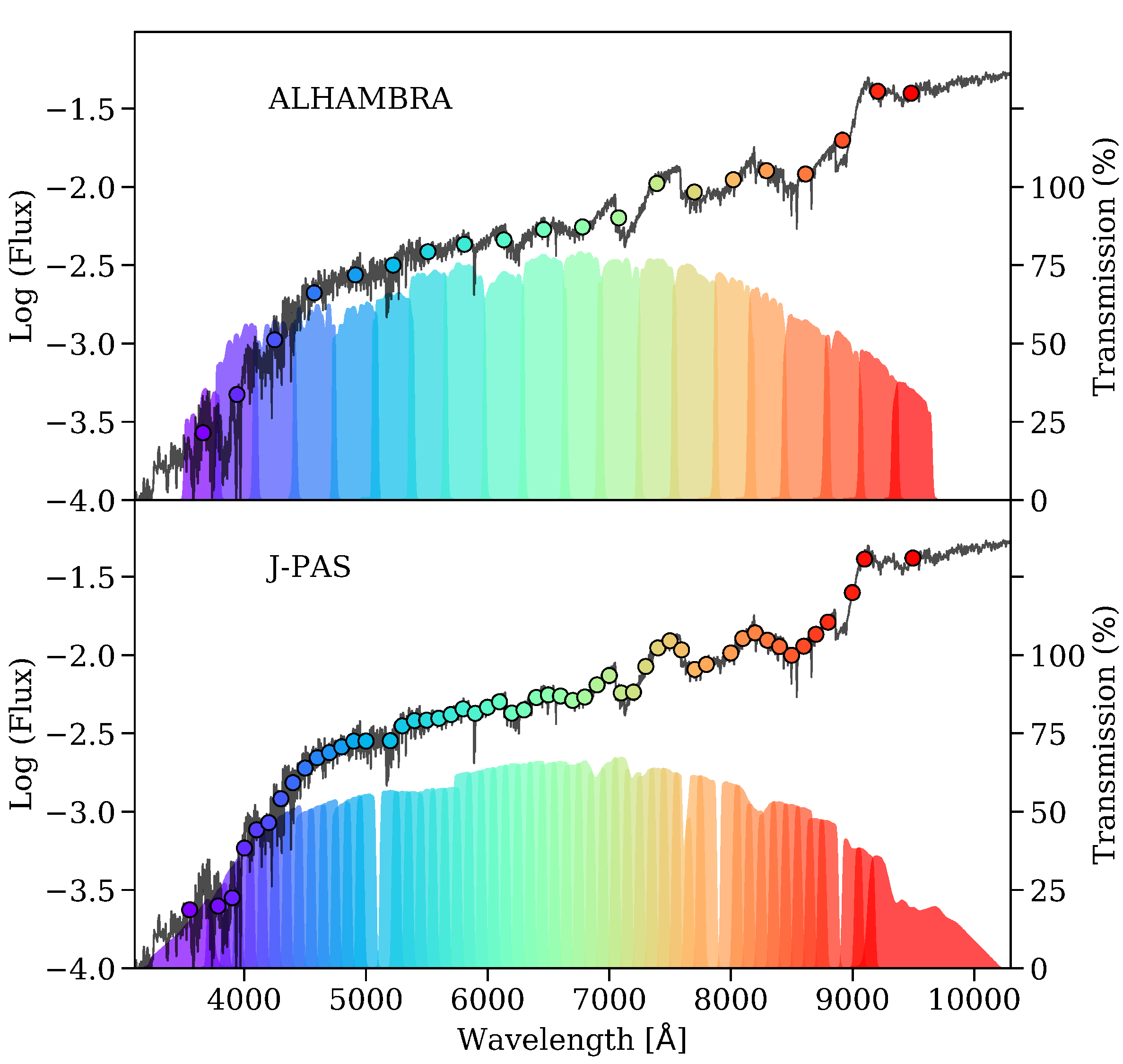}
    \caption{The top panel shows the spectral range of the ALHAMBRA survey at CALAR ALTO showing the narrow filters overplotted on a model SBF spectrum of $10$\,Gyr and solar metallicity computed with a Kroupa Universal IMF, which is representative of ETGs. The bottom panel shows the filter system of the OAJ/J-PAS survey, which is composed of $56$ narrow-band filters.}
    \label{fig:sed_sbf_surveys}%
   \end{figure*}

We envisage a rather promising potential application of the SBF spectra computed in this work for galaxy surveys that are based on narrow-band filters. As was discussed in Section~\ref{sec:photometric}, the spectroscopic magnitudes obtained from the theoretical spectra are more accurate for box-shaped transmission responses, preferably, for narrower filters. Such filter systems provide data which lies somewhere between traditional imaging and spectroscopy, and the resulting SEDs are generally characterized by a high photometric precision. 

To illustrate such applications we show in Fig.~\ref{fig:sed_sbf_surveys} a representative theoretical SBF SED obtained with the narrow filter systems of the Advanced Large, Homogeneous Area Medium Band Redshift Astronomical Survey (ALHAMBRA) at CALAR ALTO observatory \citep{Moles08} and the Javalambre Physics of the Accelerating Universe Astrophysical Survey (J-PAS: \citealt{Benitez09,Cenarro10}). The ALHAMBRA survey is primarily intended for cosmic evolution studies covering a large-area with 20 contiguous, equal width, medium band optical filters from 3500 A to 9700\,\AA, and three standard broad bands, JHK, in the NIR \citep{Aparicio-Villegas10}. The J-PAS survey is based on 56 narrow-band ($\sim100$\,\AA) contiguous optical filters ($\lambda\lambda \sim$3500 -- 9500\,\AA) and will be soon providing low resolution (R$\sim$30) spectro-photometric data for approximately one hundred million galaxies.

The SBF SEDs in Fig.~\ref{fig:sed_sbf_surveys} show that these filters not only provide valuable information related to the overall shape of the SED, but have also the ability of measuring the strengths of prominent molecular absorption bands, particularly for the J-PAS survey. This warrants further investigations in order to untangle the information encoded in these features to constrain relevant stellar population parameters. Indeed, given the fact that the SBFs are contributed by the most luminous evolved cool stars, which dominate the light of the mean stellar population SEDs in the near-IR wavelengths, the constraints derived from these stars should also be present in the optical spectral range of the SBF SEDs. Thus, the joint study of the mean and SBF SEDs in the optical range allows us to also include constraints present in the near-IR range of the mean stellar population SED.

\section{Summary} 
\label{sec:summary}

Detailed stellar population studies of galaxies are based on fitting their observed spectra and absorption line-strength indices with the aid of theoretical model predictions. The most simple modelling approach is to provide single-age, single-metallicity stellar populations, SSPs, which can be understood as  probability distributions that are mainly characterized by  mean and variance spectra. In the standard analysis the variance spectra are usually neglected and, consequently, this wastes valuable information that is potentially useful to constrain the SFH. Here we present Surface Brightness Fluctuation spectra computed with the E-MILES stellar population synthesis models. The models cover the spectral range $\lambda\lambda$ $1680-50000$\,\AA\ at moderately high resolution, all based on extensive empirical, rather than theoretical, stellar spectral libraries. The models span the metallicity range $-2.3\leq\Mh\leq+0.26$ for a suite of IMF types and varying slopes. 

Among the main features that distinguish the SBF spectrum from the corresponding
mean spectrum we highlight that the flux of the SBF spectrum peaks at much
redder wavelengths for old stellar populations ($J,H$ bands in the SBF spectrum
and $V,R$ bands in the mean spectrum). Another remarkable feature of the SBF
spectrum is that it shows much stronger molecular band absorption, which can
erase to a great extent prominent features in the mean spectrum such as the
Ca{\sc{II}} triplet in the $I$ band. We have characterized the impact of the
stellar evolutionary models that feed our stellar population models on the
resulting SBF spectra. As expected, the impact on the SBF spectra is much larger
due to the significant contribution of the later evolutionary stages on the
resulting models. We have characterized the behaviour of the SBF spectra with
varying stellar population parameters, i.e., age, metallicity and IMF. We find
that the SBF spectra are mostly sensitive to  age and metallicity, providing
further constraints on the stellar content inferred from the standard analysis
based on the mean spectra. We also have studied the behaviour of the SBF spectra
with varying $\alpha$-enhancement. At constant total metallicity the abundance
enhanced SBF spectra are brigther by $0.1-0.2$\,mag in the optical range
whereas, at constant iron abundance, the $\alpha$-enhanced SBF spectra are
fainter by $0.3-0.4$\,mag.  

We discuss the possible use of the SBF spectra in some practical cases. Due
to the properties of the variance, the construction of the SBF spectrum
corresponding to a SFH has to be made by combining the variances and the mean
spectra of the various SSP components separately. These are then divided to
obtain the combined SBF spectrum. We also test several approaches to obtain
spectroscopic fluctuation magnitudes from the SBF spectra. The best
results are obtained by adopting the hypothesis of full correlation within the
filter wavelength domain. Note that such approach provides the brightest
possible magnitudes. We have compared these spectroscopic magnitudes to the
photometric ones, which are computed in the standard way on the basis of
extensive stellar photometric libraries. This comparison provided us with
offsets, due to non full correlation within the filter wavelength domain, which
can be used to correct the spectroscopic magnitudes. We tabulate these offsets
for a variety of widely used set of filters and provide a detailed recipe for
measuring the spectroscopic SBF magnitudes, provided the variance and mean spectra.

We compare our models with those of various authors, which employ stellar isochrones and stellar libraries that differ from the empirical ones that feed our models. The recently published SBF model spectra of \citet{Mitzkus18} are in reasonably good agreement with our predictions, the two based on the MILES stellar spectral library. These comparisons highlighted a striking difference that is caused by Post-AGB stars and white dwarf evolution, only included in the \cite{Buzz89} models, which increase the flux of the SBF spectra blueward $\sim3000$\,\AA. Such contributions are extremely significant in the NUV range in the SBF spectra, whereas their impact on the mean spectra is only detectable in the FUV. Such sensitivity can be potentially used to study the UV upturn phenomenon present in a fraction of elliptical galaxies.

Although high quality photometric SBF magnitudes are required for a
variety of studies, we show that the spectroscopic SBF magnitudes can be also
useful at aiding and complement these studies. The SBF spectra can be used to
generate fluctuation magnitudes for a variety of situations in a rather easy
way. The SBF model spectra can be also useful to transform between varying
definitions of a given photometric filter, or to achieve precise comparisons
with fluctuation magnitudes and colours from varying sources. We compare our
spectroscopic fluctuation magnitudes with empirical SBF calibrations that are
used for distance measurements and find a reasonably good agreement.

The SBF spectra are very useful to constrain relevant stellar population parameters. In fact, by comparing the predicted SBF spectra to a representative sample of ETGs we untangle small ($<5$\%) mass-fraction contributions from extremely metal-poor ($\Mh<-1$) components. This result is particularly remarkable, given the strong insensitivity of the standard spectral analysis to these small contributions of very metal-poor stellar populations in the visible and in the near-IR spectral ranges, which are completely hidden by the dominant, old metal-rich, stellar population. Thus, SBF spectra have the potential to put very stringent constraints on the early stages of the formation of the ETGs. These results should motivate new observations and the use of catalogues of multi-wavelength SBF measurements of large sample of galaxies such as, e.g., the Next Generation Virgo Cluster Survey \citep{Cantiello18}.

Finally, we also show that the new SBF models represent an excellent opportunity  for exploiting ongoing photometric surveys, particularly those based on narrow-band filters. However, significant modelling work is urgently required to identify and characterize those spectro-photometric features that provide the most relevant constraints. 
These new models can be downloaded from the MILES website (http://miles.iac.es/).

\section*{Acknowledgments} 

We thank M. Mitzkus and C.~J. Walcher for kindly providing us with a set of
their recently published SBF model spectra. We are very grateful to the referee
who provided a very constructive and detailed comments and suggestions that
helped us to improve significantly the original draft. We thank P. Rodr{\'{\i}}guez-Beltr\'an and J. Falc\' on-Barroso for preparing the model distribution for the miles website. We are grateful to I.
Trujillo and G. van de Venn for very useful discussions. AV thanks the ESO
visitor program support (2018). This work has been supported by the grants
AYA2016-77237-C3-1-P, AYA2017-88007-C3-1-P,  AYA2015-68012-C2-01  and
MDM-2017-0737 (Unidad de Excelencia Mar{\'{\i}}a de Maeztu CAB) from the Spanish
Ministry of Science, Innovation and Universities (MCIU). This work has been
supported through the regional budget of the Consejer\'\i a de Econom\'\i a,
Industria, Comercio y Conocimiento of the Canary Islands Autonomous Community.
MB gratefully acknowledges support from the Severo Ochoa excellence programme
(SEV-2015-0548). IMN acknowledges support from the Marie Sk\l odowska-Curie
Individual {\it SPanD} Fellowship 702607.

\bibliographystyle{mnras}
\bibliography{papers}

\appendix
\section{Spectroscopic SBF magnitudes and correction factors}
\label{sec:filter_corr}

Users of the SBF spectra could be tempted to perform a direct integration for
computing magnitudes or spectral indices. However such an integration, although 
providing the correct units, would not be able to recover the involved variance
properties. Written explicitly, and taking into account the quadratic term in
the variance, we obtain the following expressions that are not equivalent

\begin{eqnarray}
    \int_{m_{\rm l}}^{m_{\rm t}} \left(\int_\lambda L_\lambda (m,t,[\mathrm{Fe/H}])\, T(\lambda)  
    \;\mathrm{d}\lambda  \right)^2 N_\Phi(m,t)\;\mathrm{d}m \neq \nonumber \\
    \int_\lambda \left( \int_{m_{\rm l}}^{m_{\rm t}} L_\lambda^2 (m,t,[\mathrm{Fe/H}])  
    N_\Phi(m,t)\;\mathrm{d}m \right)  \, T(\lambda)\;\mathrm{d}\lambda
\end{eqnarray}

\noindent that is, any property integrated over the wavelength domain will lead
to a different variance when the integration is performed either before or after
the computation of the stellar population. It so happens that the different
wavelengths in a stellar spectrum are related to each other, and such
correlations should be preserved (and propagated) in the computation of the
spectrum of a stellar population. This is the case when convolving 
the SBF spectrum with a given filter response to obtain spectroscopic photometry, or simply, when re-sampling the SBF spectrum to a different resolution.

A proper treatment requires obtaining the correlation matrix between all wavelengths during the computation of the SBF spectrum of the SSP:

\begin{equation}
    \int_{m_{\rm l}}^{m_{\rm t}}  L_{\lambda_i}(m,t,[\mathrm{Fe/H}])\,L_{\lambda_j}(m,t,[\mathrm{Fe/H}]) \; N_\Phi(m,t)\;\mathrm{d}m \;\; \forall i,j
\end{equation}

\noindent for all the wavelengths in the spectra (with the corresponding subtraction of the product $L^\mathrm{SSPmean}_{\lambda_i} \, L^\mathrm{SSPmean}_{\lambda_j}$ to transform it to a covariance) to obtain the corresponding covariance matrix. However in a practical use of the SBF spectrum such a treatment would be cumbersome. Therefore, we have explored various alternative approximations to simplify such  complexities. The most simple case is obtained under the assumption of fully correlated spectra, as discussed in the main text (see Section~\ref{sec:photometric}). In this section we show that such an approach provides the best simple solution for obtaining the spectroscopic SBF photometry, despite its limitations, which include possible issues related to the loss of full correlation in the strongest absorption lines in the SBF spectrum.

We employ here the {\it SED@} models of \cite{CLCL06}, with BaSTI 
and Padova00
isochrones and the low resolution BaSeL3.1 
stellar spectral library of \citet{basel3}, to investigate these alternatives. We derive the spectroscopic fluctuation magnitudes from the computed spectra and we also calculate the photometric magnitudes in the standard way, i.e. integrating over the isochrone the individual star fluxes in the selected band leading to 

\begin{equation}
L_{band}^\mathrm{SSPsbf}(\mathrm{sbf_{PHOT}}) \, 
=\, {\langle L_{band}^2 \rangle \over \langle L_{band} \rangle}
\;.\label{eq:lbarphot}
\end{equation}

\noindent The models are self-consistent in the sense that the two approaches lead to the same $L_{band}^\mathrm{SSPmean}$ result. Therefore we focus on comparing the results of the various approaches for obtaining $L_\lambda^\mathrm{SSPvar}$ or $L_\lambda^\mathrm{SSPsbf}$ from their corresponding spectra. The first step consists in establishing the coefficients of integration over the filter $T_{\lambda_i}$

\begin{equation}
    L_{band} = \int T(\lambda)\, L_\lambda \;\mathrm{d}\lambda  = \sum_{i=1}^N T_{\lambda_i} L_{\lambda_i}
\end{equation}

\noindent where $L_{\lambda_i}$ is either the spectrum of an individual star or the mean SSP spectrum, which in this particular case are defined by their resolution.

Having defined $T_{\lambda_i}$ we now consider four possible alternatives. The first two cases represent the most extreme assumptions regarding the correlation between the different wavelengths, namely, full (positive) correlation ($\rho=1$, with $\rho$ representing the correlation coefficient) and null correlation ($\rho=0$), which lead to

\begin{eqnarray}
    L_{band}^\mathrm{SSPvar}(\rho=1) &=&\left( \sum_{i=1}^N T_{\lambda_i} \sqrt{L_{\lambda_i}^\mathrm{SSPvar}}\right)^2   \\
    L_{band}^\mathrm{SSPvar}(\rho=0) &=&   \sum_{i=1}^N T^2_{\lambda_i} L_{\lambda_i}^\mathrm{SSPvar} 
\end{eqnarray}

The other two cases correspond to the direct integration of the filter profile
with the SBF spectrum $L_{band}^\mathrm{SSPsbf}(\mathrm{sbf})$, or the variance
spectrum $L_{band}^\mathrm{SSPvar}(\mathrm{var})$

\begin{eqnarray}
    L_{band}^\mathrm{SSPsbf}(\mathrm{sbf}) &=&   \sum_{i=1}^N T_{\lambda_i} L_{\lambda_i}^\mathrm{SSPsbf} \\
    L_{band}^\mathrm{SSPvar}(\mathrm{var}) &=& \sum_{i=1}^N T_{\lambda_i} L_{\lambda_i}^\mathrm{SSPvar}   
\end{eqnarray}

\noindent It is worth noting that the case of
$L_{band}^\mathrm{SSPsbf}(\mathrm{sbf})$ is not a mathematically correct
approach as it does not allow us to recover the variance within the band. The
SBF obtained in this way represents the ratio between the variance and the mean
spectra within the photometric band and, consequently, it does not preserve the
properties of a variance and a {\it bona-fide} SBF. Finally, in the last case we
add the individual variances weighted by the filter transmission, which resembles
the method proposed for the computation of the variance of a Star Formation
History. Note that the SFH follows a frequency distribution and, therefore, its
use can be justified in the scaling relations of the statistical properties of
the stellar populations. However, in the present case neither the transmission
curve represents a frequency distribution in the statistical sense, nor can we
make use of any scaling relation based on the statistical properties of the
spectrum.

 \begin{figure}
   \includegraphics[width=0.49\textwidth]{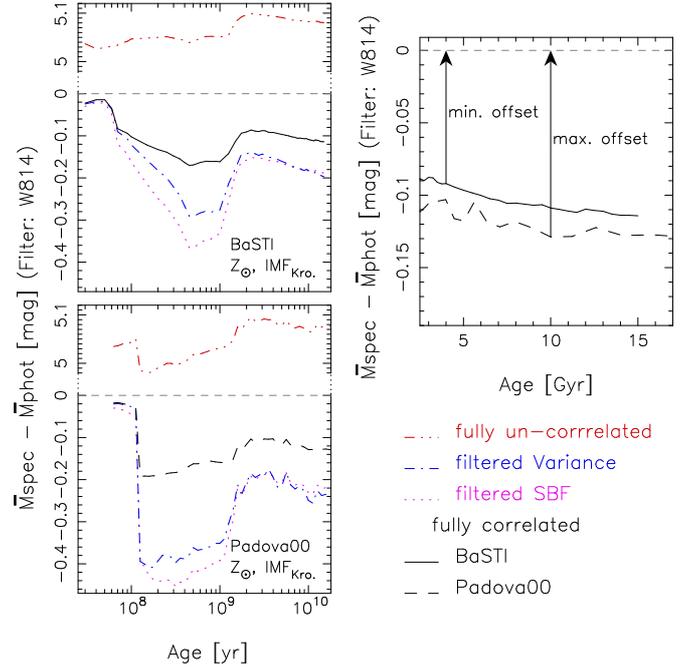}
    \caption{SBF magnitude difference obtained for the F814W filter by
comparing the photometric magnitude to the one synthesized from the SBF spectrum
using different methods. The results are shown as a function of age for solar
metallicity and Kroupa Universal IMF. We selected the F814W filter as it
provides  the poorest agreement and, therefore, it helps at emphasizing the
difference between these varying methods. The left panels represent the results
obtained for the models based on the BaSTI isochrones (upper) and Padova00
(lower). Note that there is a scale change between the upper and lower part in
these two panels. The solid (top-left panel for BaSTI) and dashed (bottom-left
panel for Padova00) black lines represent the spectroscopic SBF magnitudes
assuming the hypothesis of full correlation within the filter passband
($L_{band}^\mathrm{SSPvar}(\rho=1)$ in the text). The red dotted-dotted-dashed
lines in the left panels correspond to the fully uncorrelated case
($L_{band}^\mathrm{SSPvar}(\rho=0)$ in the text). The pink dotted lines in the
same panels represents the direct application of the filter over the SBF
spectrum ($L_{band}^\mathrm{SSPsbf}(\mathrm{sbf})$). Finally the blue
dashed-dotted lines show the results for a direct application of the filter
response over the variance spectrum ($L_{band}^\mathrm{SSPvar}(\mathrm{var})$).
The black thin dashed line represents the perfect match in the left panels. The
top-right panel shows the maximum and minimum offsets obtained with the
full correlation hipothesis for the two set of isochrones. These values are the
ones tabulated in Tables ~\ref{tab:filter_maxmin_jhonson},
\ref{tab:filter_maxmin_sdss} and \ref{tab:filter_maxmin_hst}.}
    \label{fig:appendix}%
\end{figure}

We have applied these four approaches to obtain the spectroscopic SBF
magnitudes corresponding to a variety of standard filter systems and the two sets
of isochrones. We find similar trends for all the filters. We illustrate the
results in  Fig.~\ref{fig:appendix} for the F814W HST filter. This filter is the
 that which shows the largest difference between the spectroscopic and
photometric magnitudes. The figure shows the ratio of the spectroscopic
SBF magnitude obtained with different methods with respect the photometric
value and indicates that the fully uncorrelated case is completely out of
scale. It strongly underestimates  the reference photometric SBF value (note the
change of scale between the upper and lower parts of the diagram). The other
three cases produce far better results, with the fully correlated case,
$L_{band}^\mathrm{SSPvar}(\rho=1)$, the one which provides the best results.
This approach overestimates slightly the photometric SBF magnitude. We note
that, formally, this case represents the maximum allowed overestimation.
Finally, the cases in which we apply a direct integration of the filter response
over the SBF or the variance spectrum lead to slightly larger overestimates.
Their performance depends on the age range regime and the filter. Despite the
fact that these two approaches are reasonably straightforward to compute, we
discourage their use due to such dependencies on  age and filter,  and because
they do not provide  smaller deviations with respect to the photometric
magnitudes. 

Given these results we select the full correlation wavelength approach to obtain the spectroscopic magnitudes. Figure~\ref{fig:appendix} shows that this method leads to better results at lower ages, where the same massive stars dominate the luminosity in a fairly wide wavelength range. The largest differences are obtained for intermediate ages, specifically between $0.1$ and $2$\,Gyr, where AGB stars maximize their relative contribution to the total light in this spectral range. However, the fraction with which these stars contribute is a strong function of wavelength, leading to a departure from the full correlation hypothesis. Finally for the old age regime, above $2$\,Gyr the difference in magnitude does not change significantly as a function of age.

Tables~\ref{tab:filter_maxmin_jhonson}, \ref{tab:filter_maxmin_sdss} and
\ref{tab:filter_maxmin_hst} list the maximum and minimum offsets that are
required to match the spectroscopic SBF magnitudes to the photometric ones due
to non full correlation effects. These limiting values correspond to the
absolute minimum and maximum differences obtained within the age regime older
than $4$\,Gyr ($5$\,Gyr for the metallicity $Z=0.040$ for the BaSTI isochrones)
and by varying the input set of isochrones (Padova00 and BaSTI). These offsets
are provided for different metallicity bins. When a metallicity value is missing
in one of the two isochrone sets employed here, we interpolate linearly the
corresponding minimum and maximum values found for the closest bracketing
metallicities that are available. We stress here that such offsets should not be
extrapolated to lower ages, as illustrated in the right panel of
Fig.~\ref{fig:appendix}.

\begin{table*}
\centering
\caption{Maximum and minimum offsets derived for each metallicity bin for
the Galex NUV and the Johnson-Cousin broad band filters. These offsets are
meant to be applied to the values obtained with Eq.~\ref{eq:SBF_band2} to
correct the spectroscopic fluctuation magnitudes from non full correlation
effects within the filter passband. These values were derived from varying the
input set of isochrones (Padova00 and BaSTI) over the old age regime ($4 -
13$\,Gyr). The absolute minimum and maximum offsets obtained among all the
metallicity bins covered by our models are highlighted in bold face.}
\label{tab:filter_maxmin_jhonson}
\begin{tabular}{ccccccccccccc}
 Z & \multicolumn{2}{c|}{NUV$_\mathrm{GALEX}$}& \multicolumn{2}{c|}{U }& \multicolumn{2}{c|}{B2}& \multicolumn{2}{c|}{B3}& \multicolumn{2}{c|}{V }& \multicolumn{2}{c|}{R$_\mathrm{Cousin}$ }\\
 & max. & min. & max. & min. & max. & min. & max. & min. & max. & min. & max. & min. \\
\hline
 0.0001 &      -0.069 &      -0.011 &      -0.025 &{\bf -0.009}&      -0.035 &{\bf -0.014}&      -0.037 &      -0.015 &      -0.008 &{\bf -0.002}&      -0.017 &{\bf -0.001}\\
 0.0003 &      -0.073 &      -0.009 &      -0.028 &      -0.009 &      -0.041 &      -0.014 &      -0.043 &{\bf -0.014}&      -0.005 &      -0.002 &      -0.014 &      -0.001 \\
 0.0004 &      -0.075 &      -0.009 &      -0.031 &      -0.010 &      -0.047 &      -0.014 &      -0.049 &      -0.014 &      -0.005 &      -0.002 &      -0.012 &      -0.001 \\
 0.0006 &      -0.074 &      -0.008 &      -0.037 &      -0.012 &      -0.058 &      -0.016 &      -0.061 &      -0.017 &      -0.005 &      -0.002 &      -0.008 &      -0.001 \\
 0.0010 &      -0.072 &      -0.008 &      -0.038 &      -0.014 &      -0.065 &      -0.021 &      -0.069 &      -0.023 &      -0.005 &      -0.002 &      -0.008 &      -0.002 \\
 0.0020 &      -0.080 &      -0.011 &      -0.042 &      -0.018 &{\bf -0.082}&      -0.027 &{\bf -0.086}&      -0.029 &      -0.008 &      -0.004 &      -0.019 &      -0.010 \\
 0.0040 &      -0.095 &      -0.029 &      -0.031 &      -0.021 &      -0.045 &      -0.033 &      -0.048 &      -0.035 &      -0.009 &      -0.006 &      -0.050 &      -0.025 \\
 0.0080 &{\bf -0.098}&      -0.034 &      -0.028 &      -0.019 &      -0.040 &      -0.030 &      -0.043 &      -0.032 &      -0.014 &      -0.008 &      -0.092 &      -0.060 \\
 0.0100 &      -0.097 &      -0.023 &      -0.031 &      -0.017 &      -0.040 &      -0.028 &      -0.042 &      -0.030 &      -0.015 &      -0.008 &      -0.094 &      -0.066 \\
 0.0190 &      -0.041 &      -0.010 &      -0.040 &      -0.026 &      -0.039 &      -0.027 &      -0.042 &      -0.028 &{\bf -0.019}&      -0.014 &{\bf -0.102}&      -0.066 \\
 0.0240 &      -0.045 &      -0.009 &      -0.039 &      -0.025 &      -0.038 &      -0.024 &      -0.040 &      -0.026 &      -0.018 &      -0.014 &      -0.098 &      -0.071 \\
 0.0300 &      -0.025 &      -0.004 &{\bf -0.058}&      -0.029 &      -0.044 &      -0.021 &      -0.047 &      -0.022 &      -0.016 &      -0.013 &      -0.093 &      -0.061 \\
 0.0400 &      -0.020 &{\bf -0.004}&      -0.056 &      -0.027 &      -0.041 &      -0.015 &      -0.044 &      -0.016 &      -0.015 &      -0.011 &      -0.084 &      -0.060 \\
 \\
 Z & \multicolumn{2}{c|}{I$_\mathrm{Cousin}$ }& \multicolumn{2}{c|}{J$_\mathrm{Johnson}$}& \multicolumn{2}{c|}{J$_\mathrm{Bessell}$}& \multicolumn{2}{c|}{J$_\mathrm{2MASS}$}& \multicolumn{2}{c|}{H }& \multicolumn{2}{c|}{K }\\
 & max. & min. & max. & min. & max. & min. & max. & min. & max. & min. & max. & min. \\
\hline
 0.0001 &      -0.011 &{\bf 0.000}&{\bf -0.009}&{\bf 0.000}&{\bf -0.008}&{\bf 0.000}&{\bf -0.007}&{\bf 0.000}&{\bf -0.006}&{\bf 0.000}&      -0.007 &{\bf 0.000}\\
 0.0003 &      -0.008 &      0.000 &      -0.003 &      0.000 &      -0.002 &      0.000 &       -0.001 &      0.000 &      -0.002 &       0.000 &      -0.002 &      0.000 \\
 0.0004 &      -0.007 &      0.000 &      -0.003 &      -0.001 &      -0.002 &      0.000 &      -0.002 &      0.000 &      -0.002 &      0.000 &      -0.002 &      0.000 \\
 0.0006 &      -0.007 &      0.000 &      -0.004 &      -0.001 &      -0.002 &      0.000 &      -0.002 &      0.000 &      -0.002 &      0.000 &      -0.002 &      0.000 \\
 0.0010 &      -0.007 &      -0.001 &      -0.004 &      -0.002 &      -0.002 &      0.000 &      -0.002 &      0.000 &      -0.002 &      0.000 &      -0.003 &      0.000 \\
 0.0020 &      -0.019 &      -0.004 &      -0.003 &      -0.001 &      -0.002 &      0.000 &      -0.001 &      0.000 &      -0.002 &      0.000 &      -0.004 &      0.000 \\
 0.0040 &      -0.036 &      -0.010 &      -0.002 &      0.000 &      -0.001 &      0.000 &      -0.001 &      0.000 &      -0.002 &      -0.001 &      -0.003 &      0.000 \\
 0.0080 &      -0.048 &      -0.029 &      -0.002 &      0.000 &      -0.001 &      0.000 &      -0.001 &      0.000 &      -0.003 &      -0.001 &      -0.006 &      -0.001 \\
 0.0100 &      -0.051 &      -0.035 &      -0.002 &      0.000 &      -0.001 &      0.000 &      -0.001 &      0.000 &      -0.003 &      -0.001 &      -0.007 &      -0.001 \\
 0.0190 &      -0.080 &      -0.051 &      -0.004 &      -0.001 &      -0.001 &      0.000 &      -0.001 &      0.000 &      -0.004 &      -0.001 &      -0.010 &      -0.003 \\
 0.0240 &      -0.084 &      -0.058 &      -0.004 &      -0.001 &      -0.001 &      0.000 &      -0.001 &      0.000 &      -0.004 &      -0.001 &      -0.011 &      -0.004 \\
 0.0300 &      -0.088 &      -0.068 &      -0.004 &      -0.002 &      -0.002 &      -0.001 &      -0.002 &      0.000 &      -0.005 &      -0.002 &      -0.012 &      -0.004 \\
 0.0400 &{\bf -0.095}&      -0.071 &      -0.005 &      -0.003 &      -0.002 &      -0.001 &      -0.002 &      0.000 &      -0.004 &      -0.002 &{\bf -0.014}&      -0.007 \\
\end{tabular}
\end{table*}

\begin{table*}
\centering
\caption{As Table ~\ref{tab:filter_maxmin_jhonson} but for SDSS filter responses.}
\label{tab:filter_maxmin_sdss}
\begin{tabular}{ccccccccccc}
 Z & \multicolumn{2}{c|}{u }& \multicolumn{2}{c|}{g }& \multicolumn{2}{c|}{r }& \multicolumn{2}{c|}{i }& \multicolumn{2}{c|}{z }\\
 & max. & min. & max. & min. & max. & min. & max. & min. & max. & min. \\
\hline
 0.0001 &      -0.028 &{\bf -0.011}&      -0.025 &      -0.010 &      -0.008 &{\bf 0.000}&      -0.009 &{\bf 0.000}&      -0.009 &{\bf 0.000}\\
 0.0003 &      -0.029 &      -0.012 &      -0.028 &{\bf -0.009}&      -0.004 &      0.000 &      -0.006 &      0.000 &      -0.005 &      0.000 \\
 0.0004 &      -0.030 &      -0.012 &      -0.031 &      -0.009 &      -0.004 &      0.000 &      -0.007 &      0.000 &      -0.005 &      -0.001 \\
 0.0006 &      -0.033 &      -0.013 &      -0.036 &      -0.010 &      -0.002 &      -0.001 &      -0.008 &      0.000 &      -0.005 &      -0.001 \\
 0.0010 &      -0.035 &      -0.014 &      -0.042 &      -0.013 &      -0.003 &      -0.001 &      -0.009 &      0.000 &      -0.005 &      -0.001 \\
 0.0020 &      -0.035 &      -0.016 &{\bf -0.056}&      -0.017 &      -0.006 &      -0.003 &      -0.026 &      -0.006 &      -0.006 &      -0.001 \\
 0.0040 &      -0.025 &      -0.015 &      -0.029 &      -0.021 &      -0.009 &      -0.006 &      -0.048 &      -0.017 &      -0.016 &      -0.001 \\
 0.0080 &      -0.023 &      -0.015 &      -0.028 &      -0.021 &{\bf -0.017}&      -0.008 &      -0.082 &      -0.061 &      -0.022 &      -0.006 \\
 0.0100 &      -0.025 &      -0.013 &      -0.029 &      -0.021 &      -0.016 &      -0.009 &      -0.089 &      -0.066 &      -0.024 &      -0.010 \\
 0.0190 &      -0.031 &      -0.021 &      -0.032 &      -0.023 &      -0.014 &      -0.010 &{\bf -0.124}&      -0.074 &      -0.059 &      -0.041 \\
 0.0240 &      -0.030 &      -0.020 &      -0.030 &      -0.022 &      -0.013 &      -0.010 &      -0.122 &      -0.082 &      -0.063 &      -0.048 \\
 0.0300 &      -0.049 &      -0.025 &      -0.031 &      -0.019 &      -0.011 &      -0.009 &      -0.119 &      -0.081 &      -0.070 &      -0.052 \\
 0.0400 &{\bf -0.049}&      -0.024 &      -0.029 &      -0.013 &      -0.010 &      -0.007 &      -0.114 &      -0.080 &{\bf -0.074}&      -0.057 \\
 \\
 Z & \multicolumn{2}{c|}{u'}& \multicolumn{2}{c|}{g'}& \multicolumn{2}{c|}{r'}& \multicolumn{2}{c|}{i'}& \multicolumn{2}{c|}{z'}\\
 & max. & min. & max. & min. & max. & min. & max. & min. & max. & min. \\
\hline
 0.0001 &      -0.039 &{\bf -0.016}&      -0.029 &      -0.011 &      -0.009 &{\bf 0.000}&      -0.011 &{\bf 0.000}&      -0.009 &{\bf 0.000}\\
 0.0003 &      -0.040 &      -0.016 &      -0.032 &{\bf -0.010}&      -0.006 &      0.000 &      -0.008 &      0.000 &      -0.003 &      0.000 \\
 0.0004 &      -0.041 &      -0.017 &      -0.035 &      -0.010 &      -0.005 &      0.000 &      -0.009 &      0.000 &      -0.003 &      0.000 \\
 0.0006 &      -0.043 &      -0.018 &      -0.041 &      -0.012 &      -0.003 &      -0.001 &      -0.010 &      0.000 &      -0.004 &      0.000 \\
 0.0010 &      -0.043 &      -0.018 &      -0.047 &      -0.014 &      -0.004 &      -0.001 &      -0.010 &      0.000 &      -0.003 &      -0.001 \\
 0.0020 &      -0.041 &      -0.020 &{\bf -0.064}&      -0.019 &      -0.008 &      -0.003 &      -0.026 &      -0.005 &      -0.003 &      -0.001 \\
 0.0040 &      -0.029 &      -0.018 &      -0.034 &      -0.025 &      -0.014 &      -0.009 &      -0.046 &      -0.016 &      -0.009 &      -0.001 \\
 0.0080 &      -0.027 &      -0.018 &      -0.032 &      -0.025 &{\bf -0.025}&      -0.012 &      -0.069 &      -0.051 &      -0.013 &      -0.004 \\
 0.0100 &      -0.029 &      -0.017 &      -0.032 &      -0.025 &      -0.024 &      -0.013 &      -0.075 &      -0.055 &      -0.014 &      -0.006 \\
 0.0190 &      -0.041 &      -0.029 &      -0.035 &      -0.026 &      -0.019 &      -0.015 &{\bf -0.104}&      -0.066 &      -0.037 &      -0.027 \\
 0.0240 &      -0.040 &      -0.029 &      -0.033 &      -0.024 &      -0.018 &      -0.015 &      -0.104 &      -0.073 &      -0.038 &      -0.029 \\
 0.0300 &      -0.075 &      -0.036 &      -0.033 &      -0.020 &      -0.016 &      -0.012 &      -0.104 &      -0.076 &{\bf -0.043}&      -0.031 \\
 0.0400 &{\bf -0.075}&      -0.036 &      -0.030 &      -0.014 &      -0.013 &      -0.011 &      -0.103 &      -0.077 &      -0.043 &      -0.034 \\
\end{tabular}
\end{table*}

\begin{table*}
\centering
\caption{As Table ~\ref{tab:filter_maxmin_jhonson} but for HST filter responses.}
\label{tab:filter_maxmin_hst}
\begin{tabular}{ccccccccc}
 Z & \multicolumn{2}{c|}{F439W }& \multicolumn{2}{c|}{F555W }& \multicolumn{2}{c|}{FW675W}& \multicolumn{2}{c|}{F814W }\\
 & max. & min. & max. & min. & max. & min. & max. & min. \\
\hline
 0.0001 &      -0.018 &{\bf -0.007}&      -0.014 &{\bf -0.004}&      -0.009 &{\bf 0.000}&      -0.017 &{\bf 0.000}\\
 0.0003 &      -0.021 &      -0.007 &      -0.013 &      -0.004 &      -0.004 &      0.000 &      -0.014 &      0.000 \\
 0.0004 &      -0.025 &      -0.007 &      -0.012 &      -0.004 &      -0.005 &      0.000 &      -0.014 &      0.000 \\
 0.0006 &      -0.032 &      -0.008 &      -0.012 &      -0.004 &      -0.005 &      0.000 &      -0.015 &      0.000 \\
 0.0010 &      -0.036 &      -0.012 &      -0.013 &      -0.005 &      -0.005 &      0.000 &      -0.016 &      -0.001 \\
 0.0020 &{\bf -0.044}&      -0.016 &      -0.019 &      -0.008 &      -0.012 &      -0.004 &      -0.034 &      -0.007 \\
 0.0040 &      -0.029 &      -0.021 &      -0.015 &      -0.012 &      -0.023 &      -0.011 &      -0.059 &      -0.020 \\
 0.0080 &      -0.025 &      -0.019 &      -0.020 &      -0.014 &{\bf -0.047}&      -0.023 &      -0.080 &      -0.056 \\
 0.0100 &      -0.026 &      -0.018 &      -0.021 &      -0.014 &      -0.046 &      -0.026 &      -0.087 &      -0.065 \\
 0.0190 &      -0.027 &      -0.019 &{\bf -0.022}&      -0.017 &      -0.042 &      -0.029 &      -0.129 &      -0.091 \\
 0.0240 &      -0.026 &      -0.018 &      -0.020 &      -0.017 &      -0.039 &      -0.030 &      -0.137 &      -0.102 \\
 0.0300 &      -0.038 &      -0.015 &      -0.019 &      -0.014 &      -0.036 &      -0.025 &      -0.147 &      -0.116 \\
 0.0400 &      -0.036 &      -0.011 &      -0.016 &      -0.012 &      -0.030 &      -0.024 &{\bf -0.164}&      -0.123 \\
 \\
 Z & \multicolumn{2}{c|}{ WFC475 }& \multicolumn{2}{c|}{WFC606}& \multicolumn{2}{c|}{WFC814}& \multicolumn{2}{c|}{}\\
 & max. & min. & max. & min. & max. & min. &\multicolumn{2}{c}{}\\
\hline
 0.0001 &      -0.025 &{\bf -0.000} &      -0.018 &{\bf -0.003}&      -0.017 &{\bf 0.000}&\multicolumn{2}{c}{} \\
 0.0003 &      -0.027 &      -0.009 &      -0.013 &      -0.003 &      -0.013 &      0.000&\multicolumn{2}{c}{} \\
 0.0004 &      -0.030 &      -0.009 &      -0.011 &      -0.003 &      -0.014 &      -0.001&\multicolumn{2}{c}{} \\
 0.0006 &      -0.034 &      -0.010 &      -0.009 &      -0.003 &      -0.014 &      -0.001&\multicolumn{2}{c}{} \\
 0.0010 &      -0.039 &      -0.012 &      -0.009 &      -0.005 &      -0.014 &      -0.001&\multicolumn{2}{c}{} \\
 0.0020 &{\bf -0.053}&      -0.016 &      -0.016 &      -0.008 &      -0.029 &      -0.006&\multicolumn{2}{c}{} \\
 0.0040 &      -0.028 &      -0.021 &      -0.021 &      -0.016 &      -0.054 &      -0.017&\multicolumn{2}{c}{} \\
 0.0080 &      -0.028 &      -0.021 &{\bf -0.032}&      -0.020 &      -0.071 &      -0.046&\multicolumn{2}{c}{} \\
 0.0100 &      -0.029 &      -0.022 &      -0.031 &      -0.020 &      -0.074 &      -0.055&\multicolumn{2}{c}{} \\
 0.0190 &      -0.032 &      -0.024 &      -0.026 &      -0.021 &      -0.114 &      -0.081&\multicolumn{2}{c}{} \\
 0.0240 &      -0.030 &      -0.022 &      -0.024 &      -0.020 &      -0.123 &      -0.091&\multicolumn{2}{c}{} \\
 0.0300 &      -0.029 &      -0.019 &      -0.022 &      -0.016 &      -0.133 &      -0.107&\multicolumn{2}{c}{} \\
 0.0400 &      -0.027 &      -0.014 &      -0.019 &      -0.016 &{\bf -0.151}&      -0.113&\multicolumn{2}{c}{} \\
\end{tabular}
\end{table*}              

\section{Alternative routes to derive observational SBF spectra}
\label{sec:galaxy_sbf_spectra}

The use of SBF spectra to constrain relevant stellar population parameters, together with the classical approach based on mean SSP spectra, has been recently tackled by \citet{Mitzkus18}. To obtain the observational SBF spectrum these authors followed the approach of \cite{TS88}, which is based on a Fourier Transform (FT) analysis. The method is aimed at disentangling the instrumental noise and the SBF signal that is correlated with the Point Spread Function (PSF). Although this approach has proven advantages, it requires PSF-correlated data to work.
 
It is interesting to explore alternative routes to obtain the SBF spectrum from already existing observations. We mention that proposed by \cite{Buzz05}, which is based on the theoretical formulation of the SBF, i.e., without recourse to the PSF and FT analysis. In practise we only require a set of galaxy spectra that allow us to obtain a mean and a variance spectrum. The method depends, however, on the way that the observational spectra have been acquired. It is worth mentioning that it is not our intention here to derive such spectra, but to provide a general outline of the method. We are presently working on developing this approach and present the results elsewhere.

For IFU galaxy data we propose the following procedure:

\begin{enumerate}
    \item Select IFU spectra along a given galaxy isophote with good signal to noise ratio, as is usually done in the standard analysis. This choice minimizes possible variations in the number of stars as well as variations of relevant stellar population parameters. Indeed, it is important to recognise that the IMF, SFH, age, metallicity and abundance ratios, are found to vary radially within early-type galaxies (e.g., \citealt{MartinNavarro18}). In other words, it is assumed that each of the selected spectra along the isophote is a particular realization of a generic population with similar physical parameters and that each spectrum is the result of a similar number of stars. 
    \item Use the selected spectra to obtain the average (mean) galaxy isophotal spectrum, $F_\lambda^\mathrm{mean}$.
    \item Obtain the variance of the sample around the mean spectrum $F_\lambda^\mathrm{var}$. Note that it is actually the application of Eq.~\ref{eq:Fvar}, which requires $F_\lambda^\mathrm{mean}$.
    \item Obtain the final SBF galaxy spectrum by dividing $F_\lambda^\mathrm{var}$ by $F_\lambda^\mathrm{mean}$.
\end{enumerate}

This procedure requires that the noise, i.e. the contribution to the fluctuation
that is not linked to the stellar population itself, is sufficiently low such
that it does not overwhelm the SBF signal. IFU data has the advantage that it
provides the evaluation of the error in the flux at each position and wavelength
during data processing. From the theoretical spectra we find that the SBF
luminosities range from about, $1\times 10^{33}$ to $1\times 10^{29}$ erg/s/\AA\
from young to old ages, respectively. Since the fluctuations diminish with
distance, we might be able to measure deviations of about $1\times 10^{-17}$
($1\times 10^{-21}$) erg/s/\AA/cm$^{-2}$ for a young (old) stellar population
for a distance of $1$\,Mpc. These fluctuations would be $100$ times dimmer at
$10$\,Mpc, and the variance associated with the IFU data should be lower than,
at least, $10$ times the SBF fluctuation value. Depending on the observations
this method might not be a reliable option with respect to that making use of
the FT (as in \citealt{Mitzkus18}).

A second procedure is to obtain the SBF spectrum from galaxy spectroscopic
surveys. In this case the requirement is that all the selected galaxies share
very similar spectra as a result of their evolutionary properties. At a given
redshift, the spectral properties of ETGs are mostly driven by galaxy mass
(traced by their central velocity dispersion measurement). This allows us to
obtain a common mean or a stacked spectrum. In this case the inferences obtained
from the analysis of the SBF do not refer to any particular galaxy (as well as
the standard SBF does not refer to any particular pixel), but to the
characteristics of the overall sample. This type of analysis cannot be performed
using the PSF-based method, since there is no common PSF, but it requires an
exquisite treatment of the observational errors and biases in the employed data.

We further emphasize that these two approaches are based on the assumption
that the ensemble of observational spectra share similar relevant stellar
population properties, otherwise the derived properties from the SBF might be
driven by the varying populations. Another important caveat to consider here is
that in this simple approach for obtaining the observational SBF spectrum we do
not take into account any potential contributions to the variance from
instrumental noise, the presence of globular clusters, foreground stars or
background galaxies. Fortunately, the instrumental noise can be remedied to a
great extent by employing high quality galaxy spectra with relatively high
signal-to-noise.

Finally, we would like to make the point that in order to fit the observational SBF spectra, or in using them as a metrics in full spectral fitting of mean spectra, an appropriate methodology has to be developed. In fact we have shown in this work the very peculiar characteristics of the SBF spectra in comparison to the mean spectra. Specifically, the SBF spectra of SSPs cannot be combined directly to obtain the corre- sponding SBF spectrum of a stellar population characterized by a given SFH. Rather, the mean and the variance spectra must be combined separately, so that we keep the properties of the latter in the resulting stellar population. Such characteristics prevents the use of current algorithms that aim at inferring relevant stellar population properties and SFHs using a full spectrum fitting approach. These limitations were discussed by \cite{Mitzkus18}, who showed that the only way to handle an observational SBF spectrum is by using it {\it a posteriori} constraint over the best fitted solution. However, neither the theoretical nor the observational SBF spectra have been properly taken into account in full spectral fitting algorithms. The inclusion of the observational SBF spectrum in spectral fitting should in principle improve our ability to derive the best matching stellar population model. Current fitting approaches simply choose the model that minimizes the $\chi^2$, i.e. the one that provides the smallest residuals, irrespective of whether these come from observational errors or not. The time is ripe for performing a detailed study to determine an optimal methodology for incorporating SBF spectra in the analysis of stellar populations. Such improvements potentially represent a major advantage for the interpretation of incoming data of nearby stellar populations taken with the new generation of large telescopes, which provide enhanced spatial resolutions. It is our intention to address this methodology elsewhere.

\end{document}